\documentclass{aa}

\usepackage{graphicx}
\usepackage{txfonts}
\usepackage[separate-uncertainty=true]{siunitx}
\usepackage{xcolor}
\usepackage{subcaption}

\DeclareUnicodeCharacter{2212}{-}
\newcommand\Tstrut{\rule{0pt}{2.6ex}}

\usepackage{hyperref}

\begin{document}

   \title{ViCTORIA project: A pilot study of the M\,49 region in the Virgo cluster in polarisation}


   \author{A. Spasic
          \inst{1}
          \and
          A. Benati \inst{2,3}
          \and
          M. Br\"uggen \inst{1}
          \and
          H. W. Edler \inst{4}
          \and
          F. De Gasperin \inst{2}
          \and
          V. Gustafsson \inst{1}
          \and 
          P. Serra \inst{5}
          \and 
          A. Bonafede \inst{2,3}
          \and
          A. Boselli \inst{6}
          \and
          H. McCall \inst{7}
          }

   \institute{Hamburger Sternwarte, Universit\"at Hamburg, Gojenbergsweg 112, 21029, Hamburg, Germany\\
              \email{angelina.spasic@uni-hamburg.de}
         \and
             INAF - Istituto di Radioastronomia, via P. Gobetti 101, 40129, Bologna, Italy 
        \and 
        Dipartimento di Fisica e Astronomia, Università di Bologna, via Gobetti 93/2, I-40129 Bologna, Italy
        \and
        ASTRON, Netherlands Institute for Radio Astronomy, Oude Hoogeveensedijk 4, Dwingeloo, 7991PD, The Netherlands
        \and
        INAF - Osservatorio Astronomico di Cagliari, Via della Scienza 5, Selargius, 09047, Italy
        \and
        Aix Marseille Univ, CNRS, CNES, LAM, Marseille, France
        \and 
        Department of Astronomy and Astrophysics, University of Chicago, Chicago, IL 60637, USA
             }

   \date{Received ; accepted }
 
  \abstract
   {Large-scale magnetic fields permeate the Intracluster Medium (ICM) of galaxy clusters. Understanding these fields gives an insight into the origin of cosmic magnetic fields and the interaction of cluster galaxies with the ICM. Using MeerKAT observations of the Virgo cluster as part of the ViCTORIA project, we probe magnetic fields via Rotation Measure (RM) synthesis.}
   {The aim of this work is to characterise the magnetic fields in and around the infalling group M\,49 in the Virgo cluster. We will probe the interaction of M\,49 with its surrounding medium through its imprint on the polarised emission and Faraday rotation.}
   {We use L-band MeerKAT data in full polarisation calibrated with the ViCTORIA MeerKAT Survey (ViMS) pipeline, which is designed for the calibration, imaging, mosaicking and RM Synthesis of the fields in the Virgo cluster, to achieve high-fidelity reconstruction of, both, total intensity and polarised emission. We analyse the background sources of the cluster for potential local changes in the magnetic field, by calculating the RM distribution.
   }
   {In the final polarised image, consisting of four pointings of 45 min, we reach a noise level of $\SI{23.3}{\mu Jy/beam}$. We detect 101 background sources in the $\sim 2.5\, \textrm{deg}^2$ region around M\,49, corresponding to a source density of 40 sources/$\deg^2$. Across the $112\, \textrm{deg}^2$ observations of the Virgo cluster, this implies a total number of 4000 polarised sources. In M\,49 we find structured, diffuse polarised emission extending in the direction of the total intensity radio tails. A higher polarisation fraction across the upper edge of the source indicates the effect of magnetic field draping. Around M\,49 we find a larger RM scatter (at $3.8\sigma$) in the wake of M\,49 compared to the region in front of it. This can inform the modelling of the turbulence created by galactic motions.}
   {}


   \maketitle
%

\section{Introduction}
\label{sec:introduction}

The large-scale structure of the universe is organised in under-dense regions known as voids and filamentary structures, at the nodes of which galaxy clusters are located. Galaxy clusters contain up to a few thousand galaxies but most of their baryonic mass is contained in the intracluster medium (ICM): a hot, ionised plasma permeating the galaxy cluster, emitting thermal Bremsstrahlung in the X-ray band \citep{1986RvMP...58....1S}. Embedded in the ICM are large-scale magnetic fields with field strengths of $\mu$G and scales of a few to thousands of kpc \citep{donnert_2018, 2004IJMPD..13.1549G}.

The existence of these fields plays an important role in multiple processes. They provide an additional term of pressure and play a key role in astrophysical processes such as heat conduction, gas mixing and propagation of cosmic rays. Furthermore, they are important for the acceleration of particles and for the formation of diffuse large-scale radio structures, such as radio relics and halos \citep{2019SSRv..215...16V}. Despite their importance, some properties, such as their origin, their amplification and their small-scale (kpc) spatial structure are not well constrained. Probing these small scales is especially important for understanding the interaction between the ICM magnetic field and the galaxies within it.

Observationally, the main method to infer magnetic fields in galaxy clusters and radio galaxies is through synchrotron emission and Faraday rotation. The synchrotron emission of radio galaxies is produced by relativistic electrons gyrating in galactic magnetic fields, emitting in the radio regime. This emission is intrinsically linearly polarised, with an initial polarisation angle $\psi_0$. Utilizing this emission the intra-cluster magnetic fields can be constrained in two ways. On one hand the polarisation itself traces the magnetic field of the emitting source. For extended radio sources synchrotron emission from relativistic electrons in partially ordered fields produces significant linear polarisation. After correcting for Faraday rotation, the observed $E$‑vectors trace the projected magnetic field orientation in the lobes while the fractional polarisation encodes how ordered the field is on beam scales. In clusters these structures are usually shaped by interactions with the ICM, such as ram-pressure stripping and shear, which leads to characteristic features in morphology and polarisation properties.
On the other hand, Faraday rotation probes the magnetised medium along the line of sight. When polarised emission passes through the ionised and magnetised ICM, the polarisation vector is rotated by a factor proportional to the observational wavelength squared ($\lambda^2$) and a property called the rotation measure (RM). The observed polarisation angle is then given by $\psi_{\mathrm{obs}} = \psi_0 + \mathrm{RM}\lambda^2$. The RM encodes the line‑of‑sight magnetic field weighted by thermal electron density according to:
\begin{equation}
    \mathrm{RM} \propto \int_{\mathrm{los}} n_e B_{||}\, \mathrm{d}r\  \si{\mathrm{rad}\,m^{-2}},
\end{equation}
where $n_e$ is the electron number density [$\mathrm{cm}^{-3}$], $B_{||}$ is the magnetic field strength along the line of sight [$\mu$G] and d$r$ is an infinitesimal path along the line of sight in Mpc. When RM measurements from multiple background sources are combined with the X‑ray–derived gas densities, RM statistics yield estimates of cluster field strength and its fluctuation spectrum \citep{2004IJMPD..13.1549G}. The density of this so-called RM-grid is limited by the sensitivity of the radio observation \citep{2019MNRAS.490.4841L}.

Over the past decades numerous studies have used the Faraday rotation of background and embedded radio sources to determine cluster magnetic fields \citep[see][]{2003A&A...412..373V, 2004A&A...424..429M, 2006A&A...460..425G, 2008A&A...483..699G, 2008MNRAS.391..521L, 2010A&A...513A..30B, 2010A&A...514A..50G, 2010A&A...514A..71V, 2012A&A...540A..38V, 2017A&A...603A.122G, 2021MNRAS.502.2518S, 2022MNRAS.514.4969V}. These works established that the central magnetic field strength is of order a few $\mu$G decreasing towards the outskirts of the cluster with a turbulent spectrum on a range of scales. These studies have relied on a small number of radio sources. With newer, more sensitive radio surveys the increased polarised source density enables a much more detailed mapping of the magnetic field \citep{2024AJ....167..226V, 2025A&A...694A.125L, 2025hsa..conf..108A, 2025A&A...700A.139P, 2026arXiv260418338L, 2026ApJ...997..214K}, although such high density RM studies are still limited to a small number of galaxy clusters. 

As the closest galaxy cluster \citep[$\SI{16.5}{Mpc}$, ][]{2024ApJ...966..145C}, the Virgo cluster is an ideal region to combine these local measurements with large-scale structures of the intracluster magnetic field. Due to its proximity, many polarised background sources can be found within the cluster region, making it possible to resolve magnetic field structures on much smaller scales than is the case for distant clusters (at the same source density). Additionally, the cluster is in a complex dynamical state, with infalling substructures such as the M\,49 group \citep{1993A&AS...98..275B, 2014A&A...570A..69B}, allowing for the study of magnetic field amplifications in both groups as well as the cluster itself. These properties make the Virgo cluster an ideal laboratory for studying the interplay between the magnetised ICM and the galaxies residing in it.

So far the Virgo cluster has been studied in polarisation on smaller scales with a focus on the polarisation of individual radio sources within it \citep{2004AJ....127.3375V, 2006A&A...447..465C, 2012A&A...545A..69W, Vollmer2013}. High-sensitivity and wide-field observations in radio are missing due to the extremely bright radio source in the centre, M\,87 \citep{2025A&A...693A.189D}. The dynamic range limitations affect the large-scale surveys covering Virgo \citep{Condon1998TheSurvey, Intema2017TheADR1, McConnell2020TheResults, Lacy2020}. This has so far restricted our ability to obtain a comprehensive and high-resolution view of the cluster’s magnetic field structure, particularly in polarisation.

The `Virgo Cluster multi-Telescope Observations in Radio of Interacting galaxies and AGN' (ViCTORIA) project aims to overcome these limitations, providing deep multi-frequency imaging of the cluster \citep{2025A&A...693A.189D}. As the first part of the project, a 144\,MHz survey performed with LOFAR High-Band Antenna (LOFAR HBA) was published in \citet{2023Edler}. The project also includes deep MeerKAT L-band observations serving two purposes: a blind HI survey aimed at mapping seven times more galaxies than previous experiments and a continuum observation in full polarisation providing substantially improved sensitivity over existing data \citep{2025A&A...693A.189D}. In the work presented here we use the L-band observations to present a polarimetric study of the Virgo cluster region centred on the subcluster Virgo B \citep{1993A&AS...98..275B} with its brightest galaxy M\,49 \citep[d = \SI{15.8}{Mpc}, ][]{2024ApJ...966..145C}. Utilising the high sensitivity of the data, we put an emphasis on the calibration and imaging of the data using our ViCTORIA MeerKAT Survey (ViMS) pipeline---developed especially for this data---to achieve high-fidelity reconstruction of total intensity and polarised emission. With the resulting sensitivity we are able to study the interaction between the ICM and the galaxies in great detail.

Our pilot field encompasses the giant elliptical galaxy M\,49 and its surrounding group. Located at a distance of roughly $\SI{1}{Mpc}$ from the centre of the cluster, near the virial radius, it is falling onto the southern outskirts of the Virgo cluster. With a virial radius of 740\,kpc and a virial mass of $\SI{4.6e13}{M_\odot}$ \citep{2019AJ....158....6S}, the group makes up about a third of the Virgo cluster mass ($M_{200} \approx 1.4 \times 10^{14}\,M_\odot$, \cite{Urban2011}). Multiwavelength studies reveal interactions between the galaxy, its group medium and the ICM of the Virgo cluster. X-ray cavities tracing both current and past activity from its active galactic nucleus \citep{2004ApJ...613..238B, 2017ApJ...848...26G, 2024A&A...690A.195S}, as well as an X-ray bright stripped tail \citep{2019AJ....158....6S} and extended 150\,kpc long radio tails at low frequency \citep{2023Edler} reveal its interaction with the surrounding medium. HII regions show its interaction with a nearby dwarf galaxy \citep{2012A&A...543A.112A}. The M\,49 region is therefore an especially interesting region for a polarisation study of the galaxy and its interaction with the surrounding medium. 

The structure of this paper is as follows: In Section~\ref{sec:data_reduction} we give an overview of the observed data and an overview of the calibration and imaging pipeline ViMS, used to reduce the data. Section~\ref{sec:results} presents the results for both the galaxy M\,49 and the local RM grid of the ICM. In Section~\ref{sec:discussion} we discuss the implications for the interaction of the ICM with Virgo B and M\,49 and a conclusion is given in Section~\ref{sec:conclusions}.\\
Throughout this paper we adopt a flat $\Lambda$CDM cosmology with $\Omega_M = 0.3$ and $H_0 = 70~\mathrm{km/s/Mpc}$. At the distance of M\,87 \citep[$d = \SI{16.5}{Mpc}$]{2007ApJ...655..144M, 2018ApJ...856..126C, 2024ApJ...966..145C} one arcsecond corresponds to $\SI{80}{pc}$. For the convention of the spectral index $\alpha$ we use $S_\nu \propto \nu^{\alpha}$. 
\section{Data and Data reduction}
\label{sec:data_reduction}
\subsection{Observations}
\label{sec:victoria}
The MeerKAT L-band observations used in this work were collected as part of the ViCTORIA \citep[acronym;][]{2025A&A...693A.189D} project. The survey covers a total area of 112$\, \mathrm{deg}^2$ extending out to approximately $r_{200}$.  The observations were carried out in two programs: MKT22008 in the 2022/23 cycle and MKT23067 in the 2023/24 cycle (PI: de Gasperin), where the first one covered the inner region up to $\SI{1}{Mpc}$ and the second one the outer region. In total 320 pointings are arranged in a hexagonal grid with a spacing of $0.58^\circ$.
The data was acquired in full-Stokes over a frequency range of \SI{856}{MHz} to \SI{1712}{MHz} with a channel width of \SI{26.123}{kHz} and a total frequency channel number of 32000. For this continuum analysis we downloaded the data with a frequency averaging of 8, giving us a channel width of \SI{208.984}{kHz} and a total frequency channel number of 4000. Each pointing has an integration time of $43-45$\, min, spanning a $\SI{4.5}{h}$ observing run. In each observing run the telescope cycled between five target fields and the gain calibrator J$1150-0023$ a total of nine times to maximise the $uv$-coverage. Additionally, the bandpass calibrator J$1939-6342$ and the standard polarisation calibrator 3C286 were targeted in each observation.

\begin{figure}[ht]
     \centering
     \includegraphics[width=0.9\columnwidth]{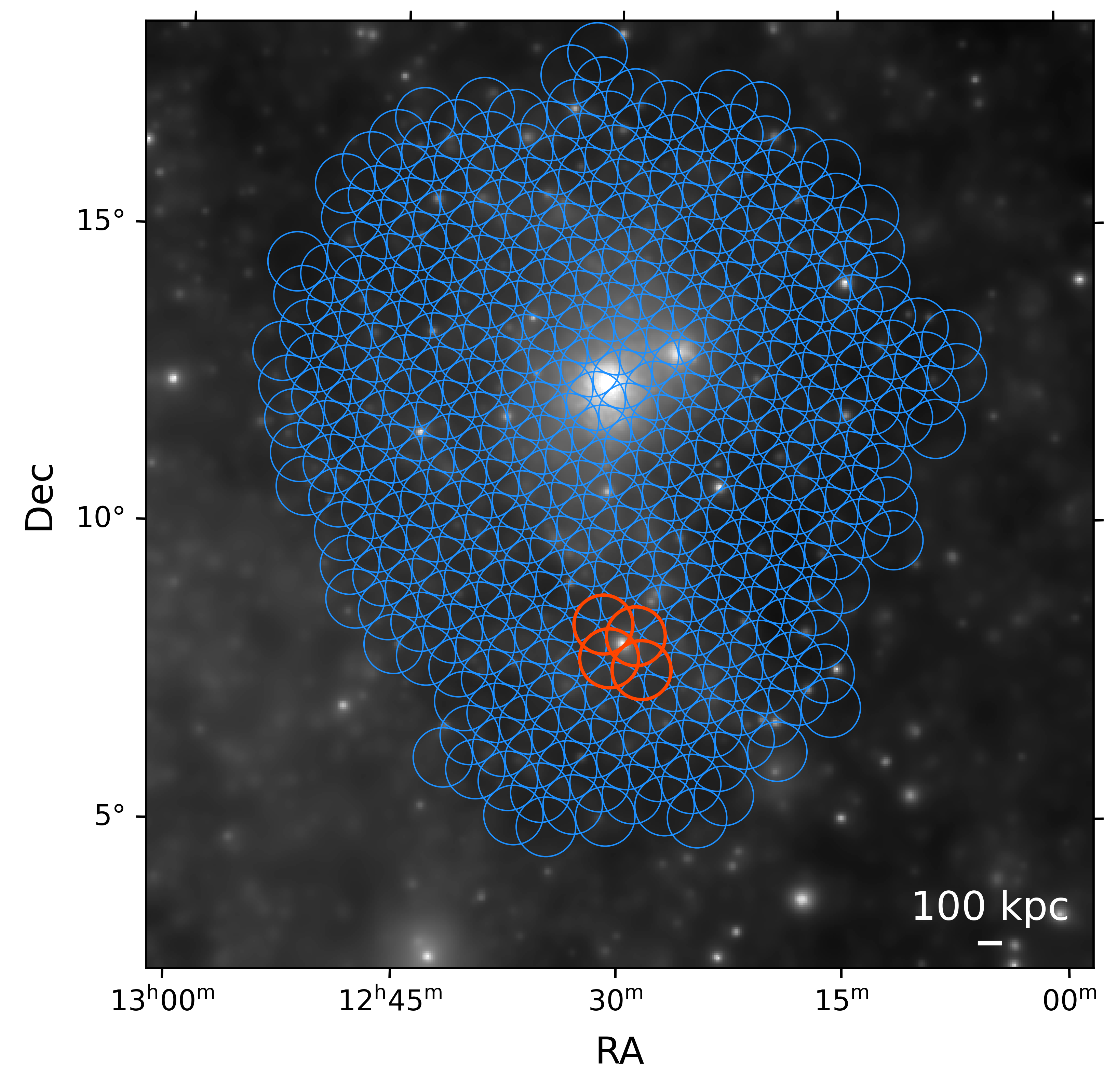}
     \caption{An overview of the MeerKAT L-band pointings over the Virgo cluster. The regions marked in red are the ones studied in this paper. The grayscale background image is the exposure-corrected and smoothed ROSAT X-ray map in the $0.1−2.4$\,keV band \citep{1994Natur.368..828B}.}
     \label{fig:pilot_field_overview}
\end{figure}

In this study, we calibrated four of those pointings across three observations in full polarisation using a custom calibration pipeline, which is explained in the following. The regions analysed here are marked in red in Fig.~\ref{fig:pilot_field_overview}.\\

\subsection{Calibration}
We perform the data reduction using the ViCTORIA MeerKAT Survey (ViMS) Pipeline \footnote{\url{https://github.com/a-benati/ViMS}}, which was designed specifically for calibrating the L-band observations of the Virgo cluster and which focuses on a careful polarisation calibration. The pipeline makes use of multiple well-established software tools for reducing the data starting from the initial flagging and continuing with the calibration, imaging, and mosaicking of the produced images. Finally, the polarisation and total intensity maps, as well as an RM map, are produced. In the following part, the steps of the pipeline are outlined with a more detailed explanation of each step and the specifics of the pipeline given in Appendix \ref{sec:pipeline_details}. 

After an initial flagging stage using \texttt{AOFlagger} \citep{2010ascl.soft10017O} and \texttt{tricolour} \citep{2022ASPC..532..541H}, we restrict the data to the range \SI{900}{MHz}−\SI{1.65}{GHz} to avoid band-pass roll-offs and average to a channel width of $\SI{1254}{\kilo \Hz}$ to reduce computational time. After applying this strategy approximately 30\% of the data in the calibrators and about 50\% of the data in the target fields is flagged. The calibration of the parallel hands (HH, VV) is performed with \texttt{CASA} \citep{2022PASP..134k4501C} using J$1939-6342$ as the bandpass and leakage calibrator and J$1150-0023$ for the gain calibration.

As the polarisation calibration is strongly affected by residual RFI, we make sure that the crosshands (HV, VH) specifically are flagged before deriving the solutions for the crosshand phase (Xf). We derive the crosshand phase and delay from the polarisation calibrator 3C286, using the full polarisation model by \cite{LinPolCal2024Memo}. To account for the ionosphere, we determine the ionospheric contribution with \texttt{Spinifex} \citep{mevius_2025_15000430} and corrupt the model accordingly using \texttt{DP3} \citep{2018ascl.soft04003V} before solving for the crosshand phase and delay.

To validate the calibration we apply the derived solutions onto the polarisation calibrator, determine its polarisation properties and compare them to the ones derived by the model, for each of the three observations calibrated for this pilot study. The results of this comparison are shown in Table~\ref{tab:pol_props_cal}. We find that the Stokes Q and U values determined for each observation are in agreement with the input model values, within a $3\sigma$ range. The uncertainties for the Stokes Q and U emission are based on the noise in the background of the image as well as an assumption of a 5\% flux scale error. The same has been done for the Stokes V observation where, for all three observations the reproduced value is close to the assumed zero circular polarisation that is assumed for the 3C286. 
The derived polarisation angle and polarisation fraction show only little variation from one observation to another. Averaged over all three, we determine a mean polarisation angle of \SI{24.4 \pm 1}{deg} and a mean polarisation fraction of $(9.6 \pm 0.2)\%$. The uncertainties for the polarisation angle have been obtained from \cite{LinPolCal2024Memo}, which includes the effect of leakage, and the ionospheric RM uncertainty. For the polarisation fraction the systematic uncertainty of the instrumental polarisation determined by \cite{2024AJ....167..273T} has been taken into account. 
The results are comparable to the polarisation angle and fraction determined from the model, which are \SI{27.7}{deg} and 9.6\% respectively. The polarisation fraction across the observations agrees with the set model within $1\sigma$. For the polarisation angle, however, the offset between the derived values and the model is approximately $3.3\sigma$ and therefore significant. We discuss this point further in Section \ref{sec:discussion}. Generally, however, these results show that the pipeline is able to reproduce the expected polarisation properties. The detailed calibration sequence and further quality checks are found in Appendix \ref{sec:pipeline_details}. 
\begin{table*}
\caption{Polarisation properties of the calibrator source 3C286 in comparison to the model data. The reference frequency is set at \SI{1.284}{GHz}.}
\label{tab:pol_props_cal}     
\centering                      
\begin{tabular}{c| c| c c c}    
\hline
\Tstrut
& Model & Obs1 & Obs2 & Obs3\\  
\hline
\Tstrut
   Stokes Q [Jy] & $0.863$ & $0.975 \pm 0.049$ & $0.971 \pm 0.049$ & $0.969 \pm 0.049$ \\
   Stokes U [Jy] & $1.249$ & $1.103 \pm 0.055 $ & $1.099 \pm 0.055$ & $1.122 \pm 0.055$ \\
   Stokes V [Jy] & $0.0$ & $0.0003 \pm 0.0004$ & $-0.0007 \pm 0.0004$ & $0.0003 \pm 0.0003$ \\
   Pol. angle [deg] & $27.7$ & $24.3 \pm 1$ & $24.3 \pm 1$ & $24.6 \pm 1$ \\
   Pol. fraction [\%] & $9.6$ & $9.6 \pm 0.2 $ & $9.5 \pm 0.2$ & $9.6 \pm 0.2$ \\
   $\mathrm{Stokes }\,V/ \mathrm{Stokes }\,I$ [\%] & $0.05$ \tablefootmark{a} & $0.002$ & $0.046$ & $0.002$\\
\hline 
\end{tabular}
\newline
\tablefoottext{a}{residual instrumental circular polarisation determined by \cite{2024AJ....167..273T}}
\end{table*}

\subsection{Self-calibration and subtraction of M\,87}
After applying the calibration solutions to the four target fields and correcting them scan-by-scan for the ionospheric RM with \texttt{Spinifex} and \texttt{DP3}, we perform four rounds of direction-independent self-calibration with \texttt{facetselfcal} \citep{2021A&A...651A.115V}. Throughout the self-calibration we restrict the solutions to scalar gains (i.e. identical solutions for the parallel hands) to preserve the relative calibration between HH and VV and therefore the amplitude of Stokes $Q$.

The central galaxy, M\,87, is very bright at radio frequencies with a flux density of $\SI{200}{Jy}$ at $\SI{1400}{\mega\Hz}$. This causes dynamic range limitations and artefacts across the images, even at angular distances of approximately two degrees from the source, when not properly removed and fields that sit in the side lobes, as it is the case for our pilot region. To mitigate this effect, we subtract a model of M\,87 whenever its phase-shifted response is detected above $5\sigma$. A detailed explanation of the \texttt{facetselfcal} settings used and the subtraction steps is found in Appendix \ref{sec:pipeline_details}.
\begin{figure*}[ht]
     \centering
     \includegraphics[scale=0.7]{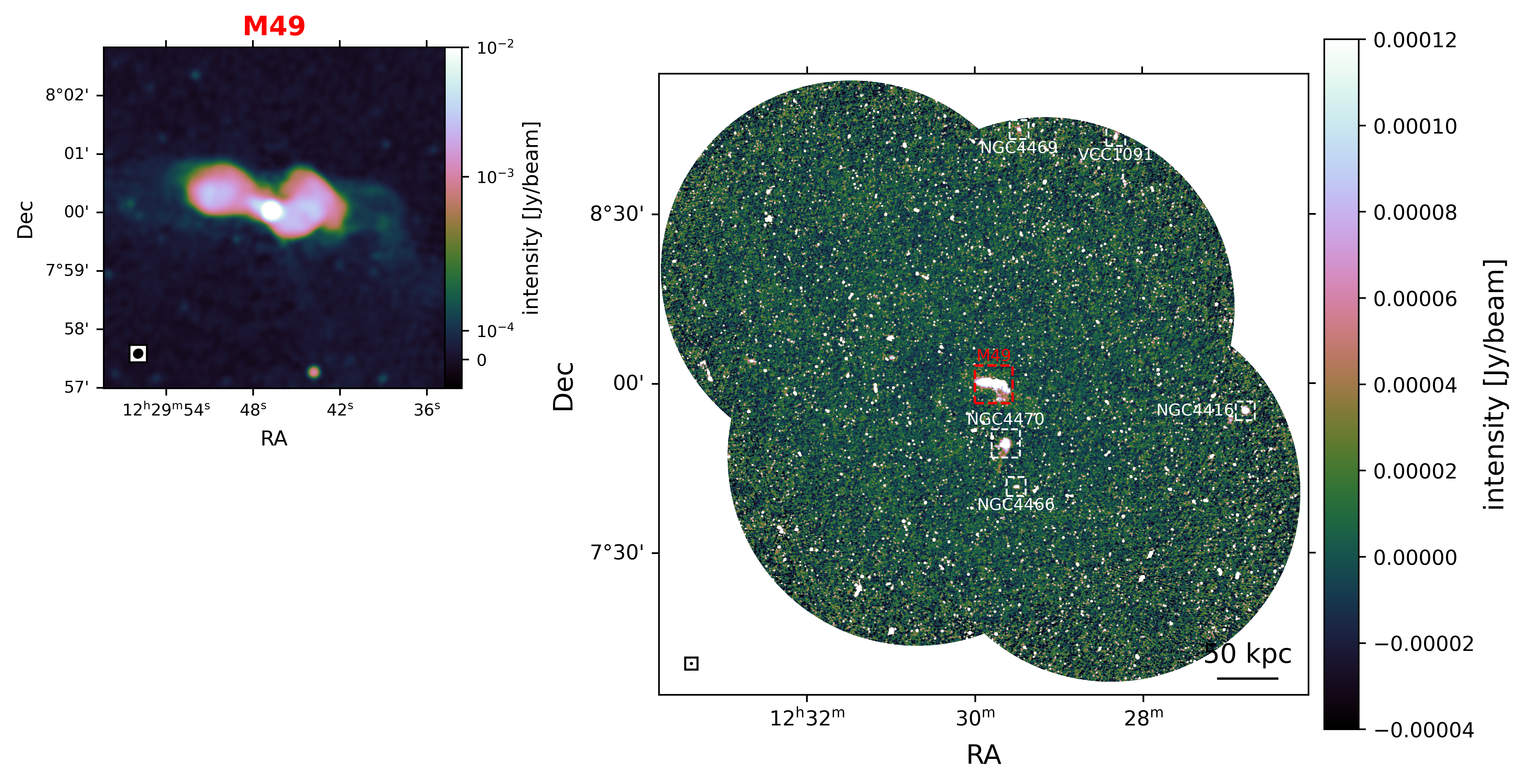}
     \caption{$\SI{1.28}{GHz}$ Stokes I image of the analysed pilot field. The size of the field is approximately $\SI{2.5}{\deg^2}$ with a sensitivity of $\SI{12.6}{\mu Jy/beam}$. The synthesised beam is $13''$. Marked with boxes are the six Virgo cluster galaxies. Marked in red is M\,49, with its zoom-in shown on the left at high resolution ($8'' \times 8''$).}
     \label{fig:StokesI_field}
\end{figure*}

\subsection{Imaging, mosaicking and RM synthesis}
\label{sec:rmsynth}
Stokes $I$ and Stokes $Q/U$ are imaged separately with \texttt{WSClean} \citep{2014MNRAS.444..606O} in order to optimise the total and polarised emission images independently from each other. To avoid bandpass depolarisation, the output channels for the polarised imaging are set to 100, such that each channel has a bandwidth of $\SI{7.5}{MHz}$. With these settings we reach a detectable Faraday depth of $>\SI{790}{rad/m^2}$. To match the resolution across the bandwidth, a taper is applied and the restoring beam is forced to be circular with a size of $13''$, enabling us to detect diffuse emission, even in polarisation, while still being able to resolve the structure within extended sources.

The per-channel images are primary beam corrected and mosaicked with \texttt{MosaicQueen} \footnote{\url{https://github.com/caracal-pipeline/MosaicQueen}}.
To reduce the effect of off-axis leakage at the edges of the Primary beam, the images are cut at 50\% FWHM of the primary beam of each individual frequency channel. The increase in leakage above $\SI{1.4}{GHz}$, as reported by \cite{LinPolCal2024Memo} is therefore mitigated by restricting the data at high frequencies to regions closer to the phase centre. In addition to that the mosaicking of the data ensures that no source is measured purely on the outer edges of an individual beam. The leakage effect in the full band mosaic is tested against the one for a cube cut at \SI{1.4}{GHz} in Appendix \ref{subsec:leakage}. The mosaicked channels in Stokes $Q$ and $U$ are then collected in cubes, excluding channels with a noise higher than $\SI{500}{\micro Jy/beam}$. This exclusion value is set to be approximately 5 times the median noise in the Stokes Q and U images across the band ($M_{\mathrm{Q/U}} \approx \SI{95}{\mu Jy}$). With this only the noisiest channels are removed. Applying this leaves 84 out of 100 channels for RM synthesis, which are weighted according to their noise during RM synthesis. 

We then employ the \texttt{RMSynth3D} routine of \texttt{RM-Tools} \citep{2026arXiv260120092V} over the range of -300 and $\SI{300}{rad/\meter^2}$ with a step size of roughly $\SI{4}{rad/\meter^2}$ with inverse-variance weighting, which yields, for our data, a Faraday depth resolution of $\SI{41}{rad/\meter^2}$ and a maximum recoverable scale of approximately $\SI{95}{rad/\meter^2}$. Using these Faraday cubes, we calculate the final polarised intensity and RM map under the assumption that all sources are Faraday simple. Additionally we deconvolve the Faraday spectra using the \texttt{RM-Tools} task \texttt{RMClean3D} for the analysis of Faraday simplicity of the sources. Full imaging parameters, noise estimates and RM synthesis steps are given in Appendix \ref{sec:pipeline_details}.

\section{Results}
\label{sec:results}

\subsection{Pilot field radio maps}
After mosaicking and imaging we produce a mean total intensity image between $\SI{900}{MHz}$ and $\SI{1.65}{GHz}$ centred on the cluster galaxy M\,49 (see Fig.~\ref{fig:StokesI_field}). The image has been mosaicked the same way as the Stokes $Q$ and $U$ images and is therefore cut at a primary beam FWHM of 50\% as well. It covers a region of roughly $\SI{2.5}{\deg^2}$ with a resolution of $\SI{13}{\arcsec}$. We determine an average noise across the whole image of $\SI{12.6}{\mu Jy/beam}$. The noise increases from $\SI{7.3}{\mu Jy/beam}$ at the centre of the map to $\SI{17.7}{\mu Jy/beam}$ at the edges.

Using the source finder \texttt{pyBDSF} \citep{2015ascl.soft02007M}, we find 2096 sources within the field above a detection threshold of 3$\sigma$. Then we crossmatch the detected sources with the Extended Virgo Cluster Catalogue \citep[EVCC]{2014ApJS..215...22K} to differentiate between cluster galaxies and background sources. Based on the catalogue, at least 35 Virgo cluster galaxies are within the considered region. We find that at least six out of the 2096 radio sources in our image belong to the Virgo cluster (marked in Fig.~\ref{fig:StokesI_field}), while the remaining 29 galaxies are not detected above $3\sigma$ in our radio images. The five other radio detected galaxies beside M\,49 are all considered to be late type spirals as classified by \cite{2014ApJS..215...22K}, while most of the ones without detected radio emission are categorised as early type galaxies. Compared to the radio undetected late type galaxies, the detected ones are larger and higher in mass, suggesting that the radio detections are limited by stellar mass rather than reflecting a difference in environment or group membership. 

\begin{figure*}[ht]
     \centering
     \includegraphics[scale=0.7]{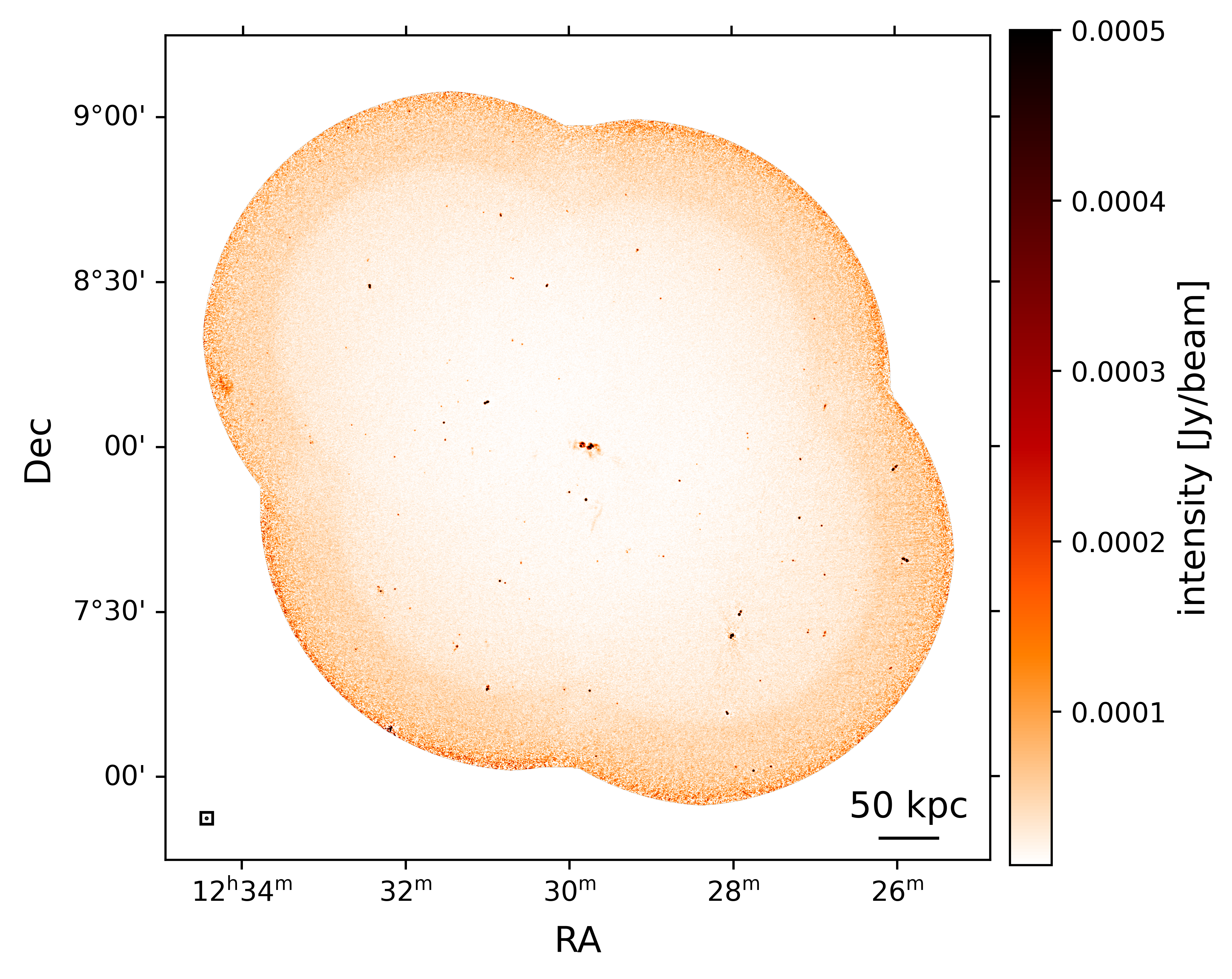}
     \caption{Ricean bias-corrected polarised intensity image of the region around M\,49 between $\SI{900}{MHz}$ and $\SI{1.65}{GHz}$. The size of the region is roughly $\SI{2.5}{\deg^2}$ with a sensitivity of $\SI{23.3}{\mu Jy/beam}$. The resolution is $13''$.}
     \label{fig:pol_field}
\end{figure*}

With the method discussed in Sec.~\ref{sec:rmsynth}, we create a polarised emission map for the regarded pilot field. This emission map is biased towards positive values due to the propagation of Stokes $Q$ and $U$. We correct this map for this Ricean bias following \cite{2012PASA...29..214G} using $\sqrt{P^2 - 2.3\sigma_{\mathrm{P}}^2}$. Here, $\sigma_{\mathrm{P}}$ is a spatial noise map derived from the dirty FDF at high Faraday depths ($|\phi| \geq \SI{100}{rad/m^2}$), as described in Appendix \ref{sec:pipeline_details}. The now Ricean bias-corrected image shown in Fig.~\ref{fig:pol_field} covers the same region as the continuum emission map and has a resolution of $13''$ as well. The sensitivity in the map varies from the centre towards the outer edges of the mosaic, from $\SI{6.6}{\mu Jy/beam}$ to $\SI{29.1}{\mu Jy/beam}$ with a mean sensitivity of $\SI{23.3}{\mu Jy/beam}$.

We determine the number of polarised sources with the help of the source finder \texttt{SoFiA} \citep{2015MNRAS.448.1922S, 2021MNRAS.506.3962W}. \texttt{SoFiA} was initially designed for HI surveys and was first employed to detect polarised sources by \cite{2025A&A...694A.125L}. We therefore take some care in constraining and masking the data before constructing a source catalogue. We use the spatial noise map described in Appendix \ref{sec:pipeline_details} as the detection threshold for \texttt{SoFiA}. We then create a detection mask by calculating the polarised intensity at each pixel, and clip it so only sources above $5\sigma$ are included. We refine the source mask thus created by requiring that each pixel is also detected in Stokes $I$ at a $3\sigma$ level. This additional masking is based on the noise map created by \texttt{WSClean}. We pass this refined mask to \texttt{SoFiA} and define sources using a merging length of 10 pixels and discarding detections with a size below 6 pixels along either the RA or Dec axis.

As a result, we find 101 polarised sources within the field, which translates to a number density of approximately 40 sources/$\deg^2$. The source density increases towards the centre of the image, where the pointings are stacked and the sensitivity is higher. Within the inner $\SI{1}{\deg^2}$ the source density is 50 sources/$\deg^2$. 

\subsection{Polarisation of M\,49}

\begin{figure}[ht]
     \centering
     \includegraphics[width=0.9\columnwidth]{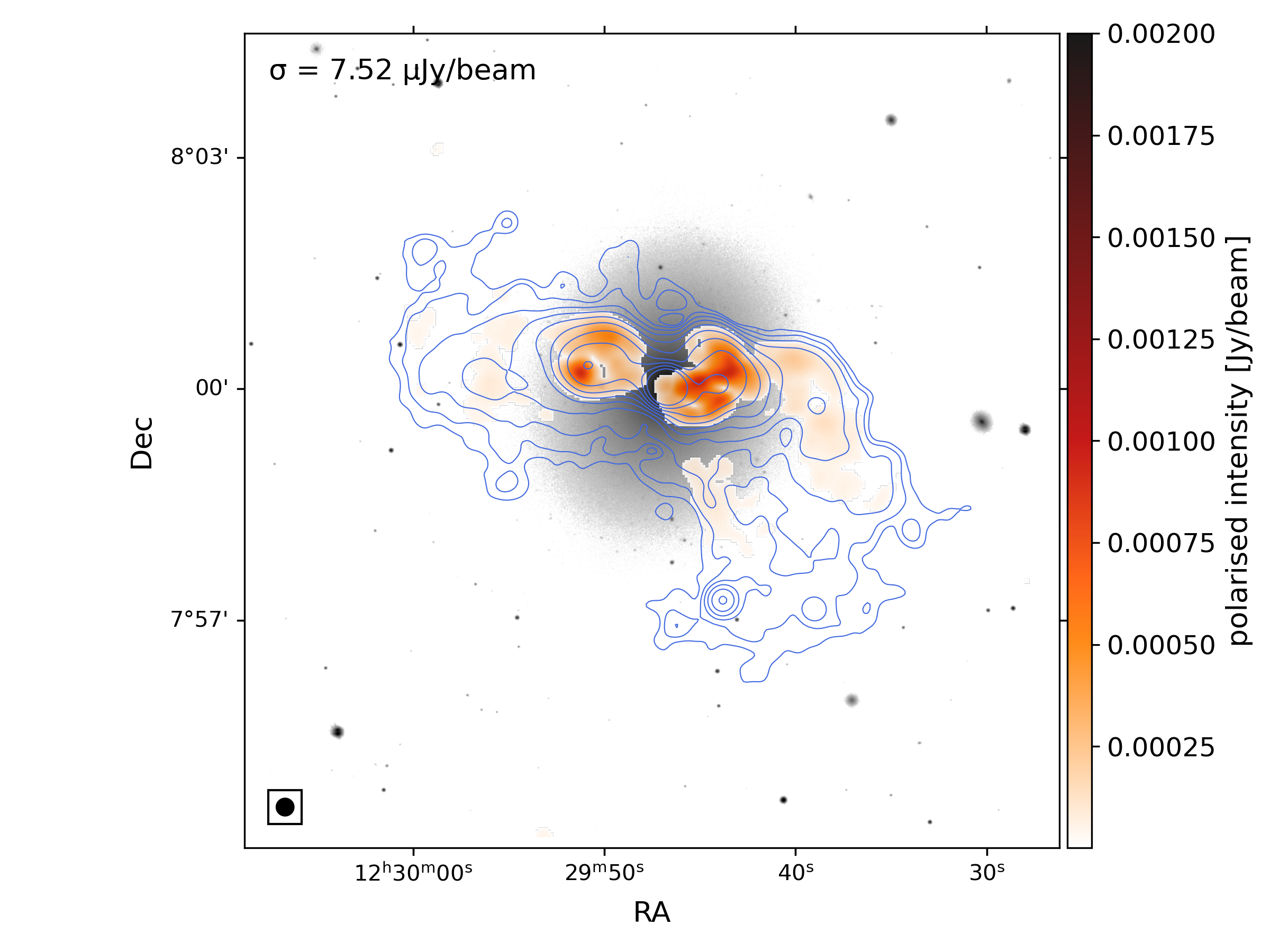}
     \caption{Bias-corrected polarised intensity map of M\,49. The blue contours show the total intensity radio emission starting at $3\sigma$ and increasing in powers of 2. The background shows the SDSS optical image in grey scales. The sensitivity of the polarised map is shown in the top left corner.}
     \label{fig:M49_opt_pol_overlay}
\end{figure}

Out of the six galaxies of our observation inside the Virgo cluster 2 galaxies are detected in polarisation, namely NGC4470 and M\,49. NGC4470 is a star forming galaxy with an observed HI tail \citep{2023A&A...676A..92B} and a prominent continuum tail in our MeerKAT observations. Along the tail we see some patches of low-level polarised emission, which will not be discussed further in this paper. M\,49 is an early-type galaxy with radio lobes that extend for roughly $10-20\,$kpc. As shown in Fig.~\ref{fig:M49_opt_pol_overlay}, the polarised emission of the source is extended following the general shape of the total emission. Especially the western radio lobe shows clear polarised emission in the outer parts, indicative of the interaction of the magnetised plasma with the surrounding medium. The inner part of the lobes shows the strongest polarised emission and shows a structure that is unlike the morphology displayed by the Stokes I emission. Compared to the simple two-sided lobes seen in Stokes I, the polarised emission seems different from the Stokes-I emission, splitting in two bright spots on each side of the radio galaxy. The central part of the source is mostly not polarised. 

\begin{figure}[ht]
     \centering
     \begin{subfigure}{\columnwidth}
     \includegraphics[width=0.9\columnwidth]{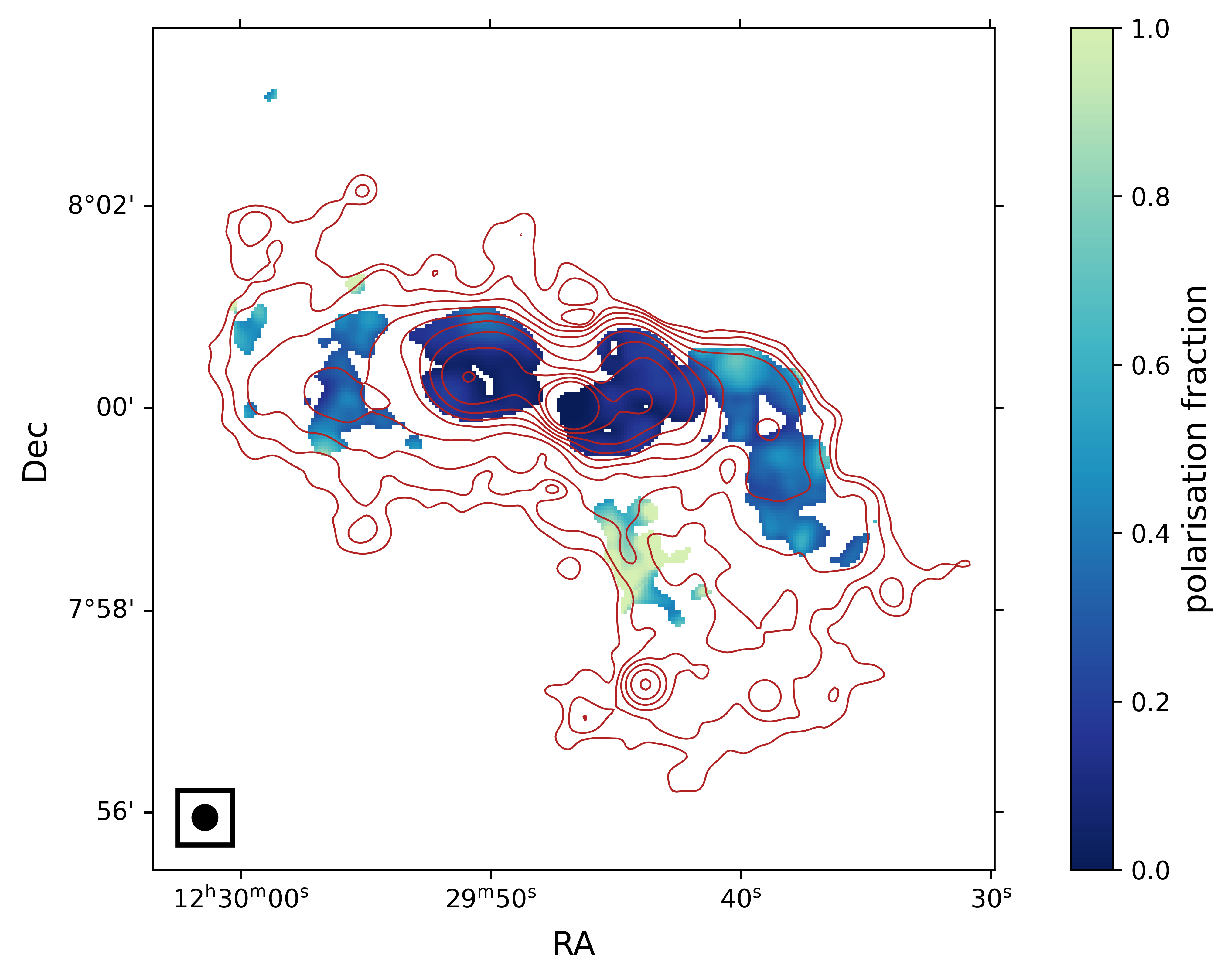}
     \caption{polarisation fraction map.}
     \label{fig:M49_polf}
     \end{subfigure}
     \begin{subfigure}{\columnwidth}
     \includegraphics[width=0.9\columnwidth]{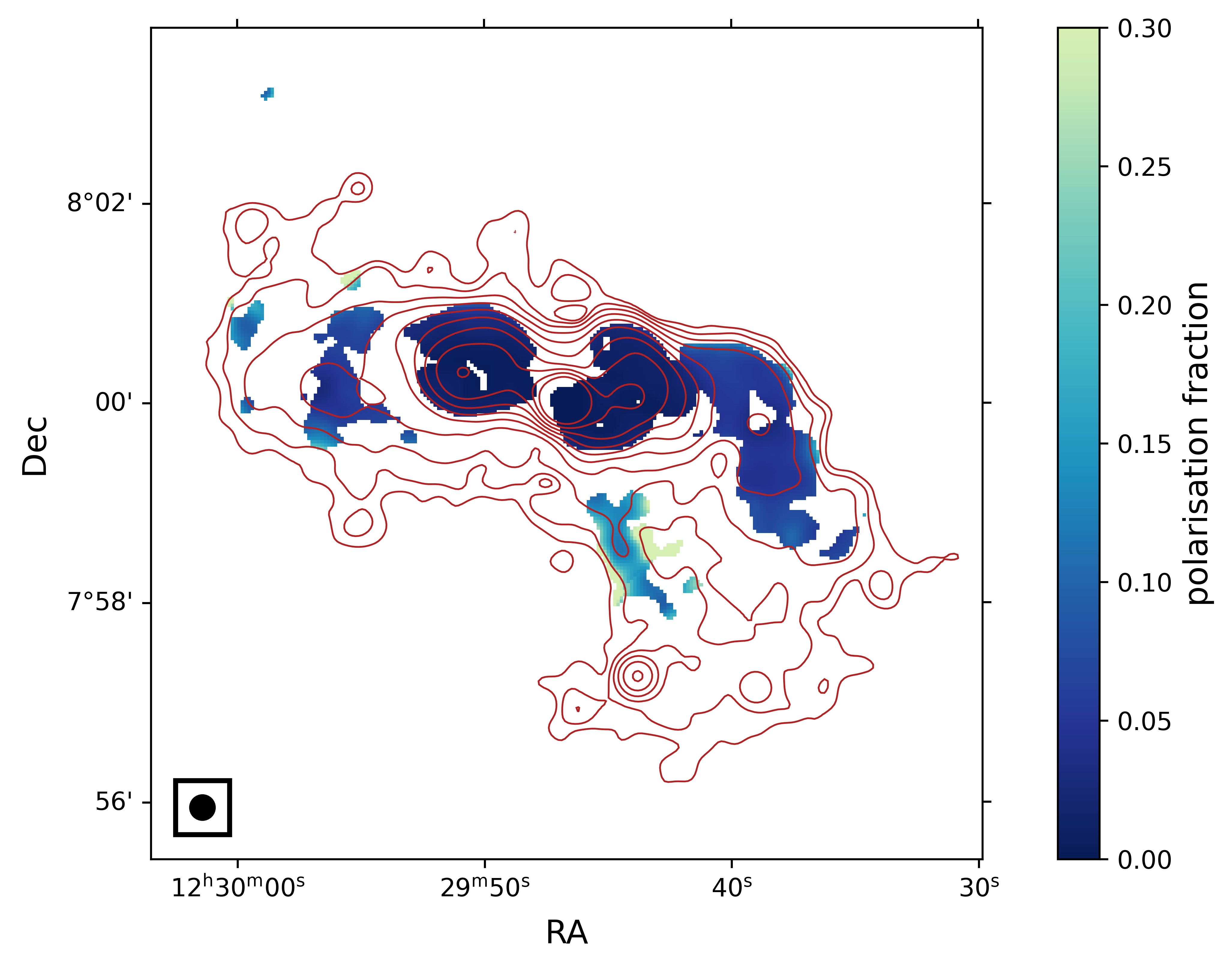}
     \caption{polarisation fraction uncertainty map. }
     \label{fig:M49_polf_err}
     \end{subfigure}
     \caption{The emission is masked at 5$\sigma$ in polarisation and $3\sigma$ in Stokes I on each individual pixel for each map. The red contours show the total intensity radio emission starting at $3\sigma$ and increasing in powers of two.}
\end{figure}

In Fig.~\ref{fig:M49_polf} we show the map of the polarisation fraction for M\,49, which we calculated from the de-biased polarisation and the total emission map via the polarisation fraction $\mathrm{polf} = \frac{P}{I}$. For the uncertainty of the polarisation fraction we assume a flux-scale error of 5\% for both the total intensity and the polarisation map and propagate it with the noise per pixel. The flux scale uncertainty had no impact on the ratio $P/I$ because it is a systematic that would affect both quantities the same way. For M\,49 we can see a relatively high polarisation fraction with a mean of $(29 \pm 5)\%$, reaching up to $(92 \pm 17)\%$ as the highest values in the extended emission towards the south. Here the emission seems to be drastically more polarised than in the central regions of the source. However, the uncertainty of that polarisation fraction is quite high as well, owing to the combination of low continuum emission and strong polarised emission. As seen in Fig.~\ref{fig:M49_polf_err}, the region towards the south has higher uncertainties than its surrounding regions.

The outer radio arms, especially in the western arm seem to show parts with stronger fractional polarisation. These spots seem to coincide with the onset of the bent arms. Further along the bent radio tails the polarisation fraction stays consistently high with values around $(35 \pm 6)\%$.

\begin{figure}[ht]
     \centering
     \includegraphics[width=0.9\columnwidth]{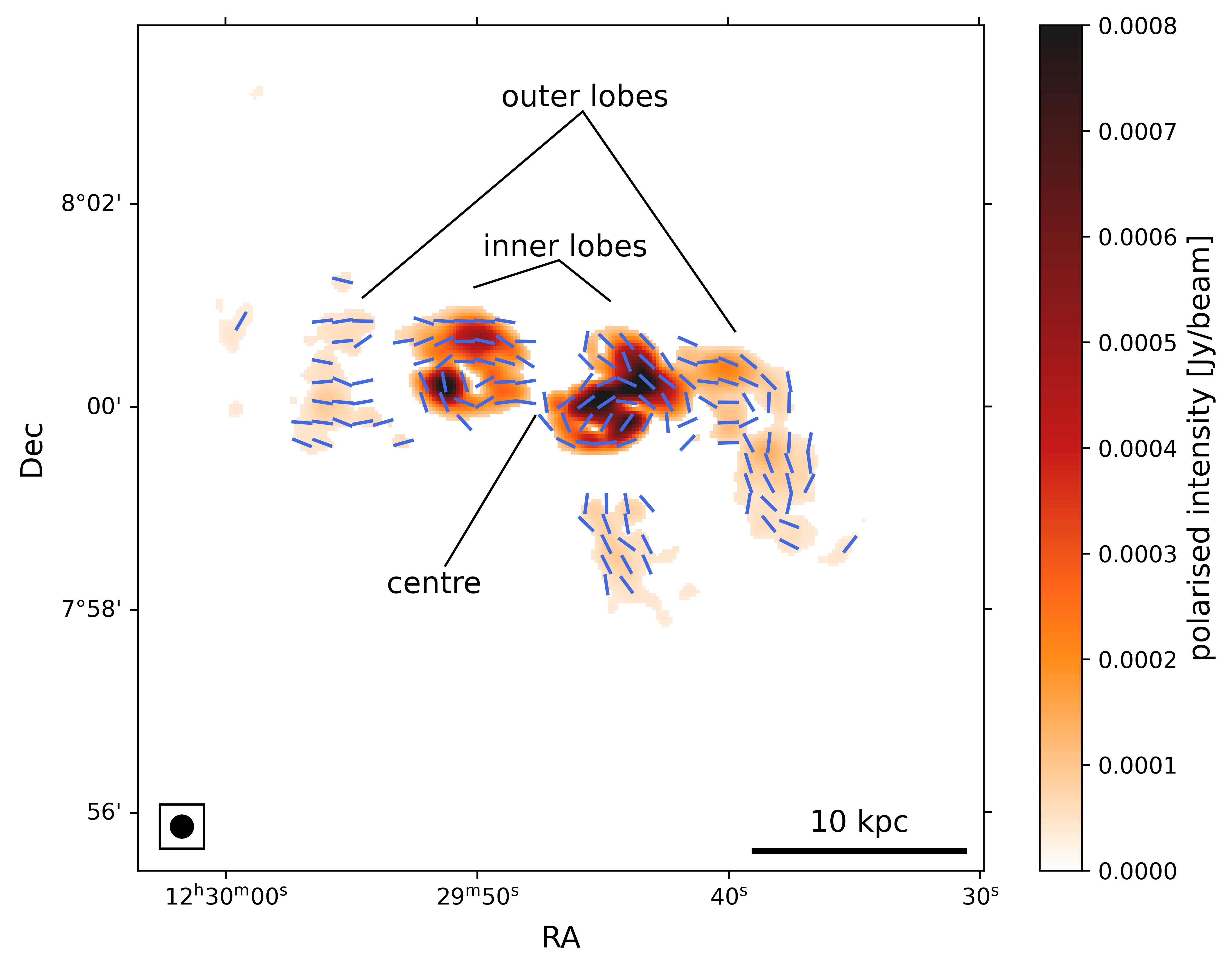}
     \caption{Polarised intensity image of the M\,49, masked for emission below $3\sigma$ in total and below $5\sigma$ in polarised emission. The lines on top of the image show the direction of the B-vectors of the source.}
     \label{fig:M49_MF}
\end{figure}

Another way to probe the polarisation properties of M\,49 is by looking at the polarisation angle of the source. Their orientation indicates the direction of the electric field. By rotating them orthogonally to their initial direction, we get a simple estimate for the B-field direction of the source. These B-vectors are plotted on top of the polarised emission in Fig.~\ref{fig:M49_MF}. In the plot the difference between the eastern and the western arm is reflected in the B-field vectors. In the western arm the vectors follow the polarised emission closely tracing the bending along the extended polarised emission up to even the slight upwards turn at the faintest emission. In the eastern arm, on the other hand, the B-vectors seem to align orthogonally to the emission direction and do not seem to follow the bending of the shape at all. There seems to be no direct correlation. The vectors within the lobes follow the structure of the polarised emission and seem to circle around the regions of lower polarised emission in the inner lobes.

\begin{figure}[ht]
     \centering
     \includegraphics[width=0.9\columnwidth]{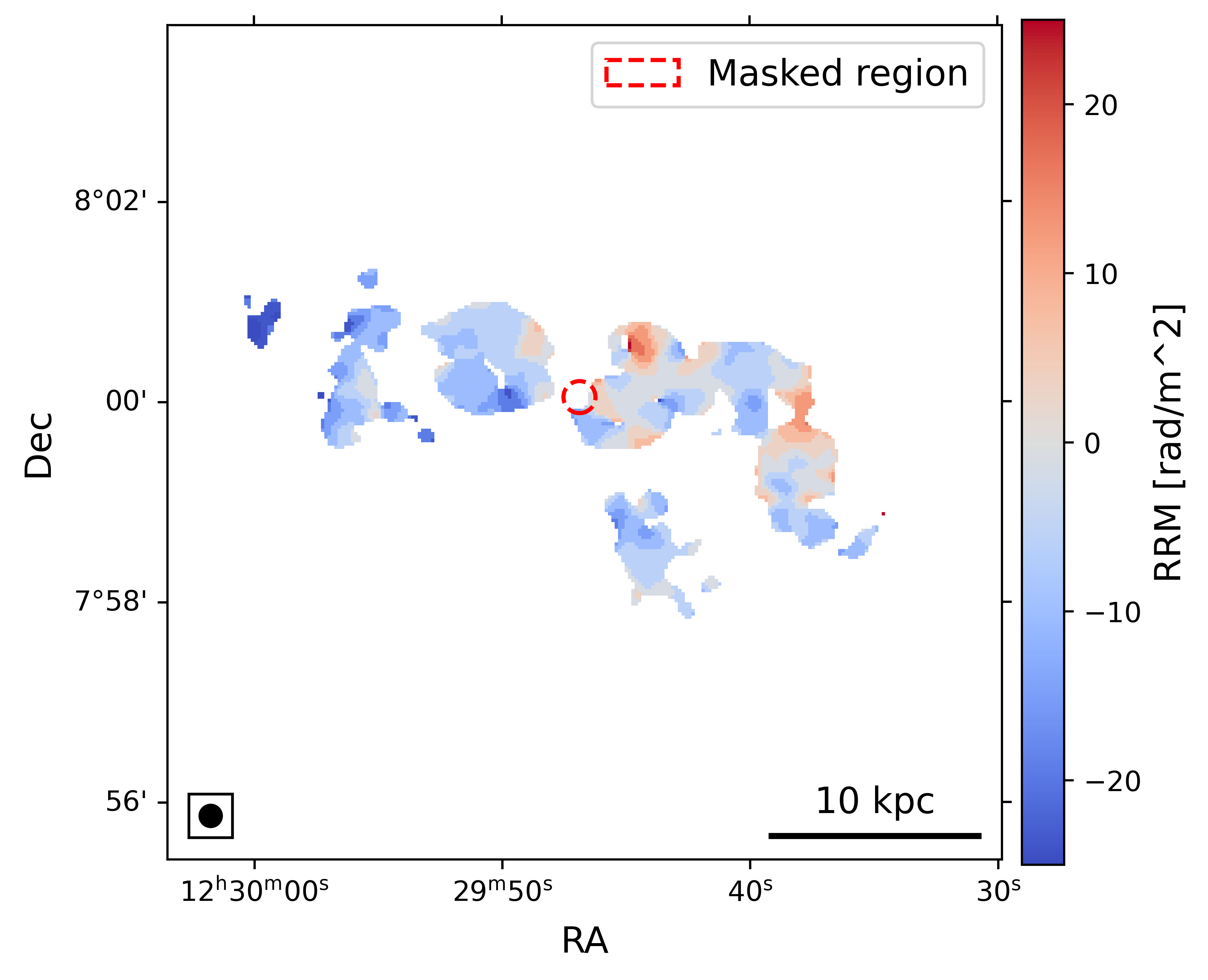}
     \caption{Galactic RM subtracted Rotation Measure map of M\,49. The central region of the sources has been masked as it does not show a Faraday simple spectrum.}
     \label{fig:M49_RM}
\end{figure}

\begin{figure}[ht]
     \centering 
     \begin{subfigure}{\columnwidth}
     \includegraphics[width=0.9\columnwidth]{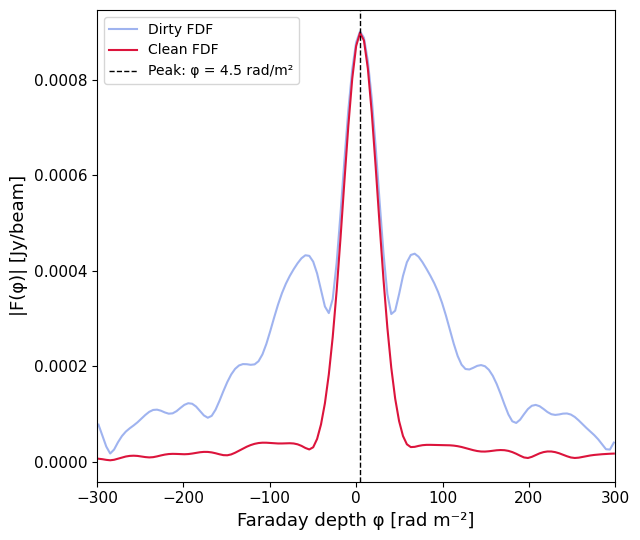}
     \caption{FDF of lobe of M\,49.}
     \label{fig:M49_FDF_lobe}
     \end{subfigure}
     \begin{subfigure}{\columnwidth}
     \includegraphics[width=0.9\columnwidth]{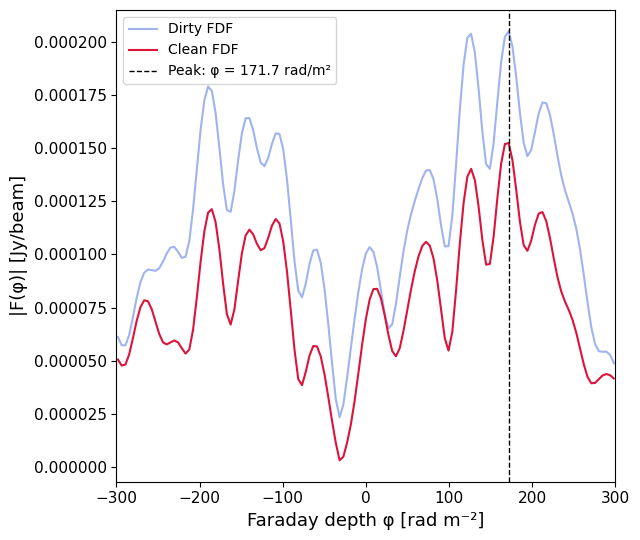}
     \caption{FDF of core of M\,49.}
     \label{fig:M49_FDF_core}
     \end{subfigure}
     \caption{Faraday spectrum for M\,49 (lobe and core). In blue the dirty spectrum is shown in comparison to the cleaned spectrum in red. The highest peak is marked with a black dotted line.}
     \label{fig:M49_FDF}
\end{figure}

The Rotation Measure map of M\,49 was corrected for the galactic RM contribution, which will be explained in the following section. Looking at the RM map of M\,49 (see Fig.~\ref{fig:M49_RM}), we see a distribution of RM ranging from $-25$ to $\SI{+25}{rad/\meter^2}$ within the lobes. In contrast to the other observed differences between the eastern and western lobe, no significant difference can be seen in RM. The eastern lobe shows slightly more negative values with a mean of $\SI{-8}{rad/\meter^2}$ compared to the western lobe which has a mean of $\SI{-3}{rad/\meter^2}$. However, these differences are well within the standard deviation for each of the two which are $\SI{6}{rad/\meter^2}$ and $\SI{7}{rad/\meter^2}$ respectively. The Faraday spectrum of the lobe (see Fig.~\ref{fig:M49_FDF_lobe}) shows a single Faraday peak, indicating that the emission is Faraday simple in M\,49.  

By contrast, the central region (masked in Fig.~\ref{fig:M49_RM}) exhibits a jump in RM with values around $\SI{130}{rad/m^2}$. These values are most likely not part of the physical properties of the source, but dominated by noise. The Faraday spectrum has no single peak but broad noise-like structures (see Fig.~\ref{fig:M49_FDF_core}). 
It is relatively bright in total radio emission with a peak flux of $\SI{0.12}{Jy/beam}$. The instrumental leakage from Stokes I into Stokes Q and U, albeit less than 2\% \citep{LinPolCal2024Memo}, can still produce a measurable contribution to the polarised emission for strong emission, such as in the core of M\,49. Additionally, M\,49 is strongly depolarised in the core region, exhibiting almost no intrinsic polarised emission. So, even after cutting the noise at 5$\sigma$, no simple Faraday peak emerges, and we can only fit the highest noise peak.

\subsection{RM distribution}
\begin{figure}[ht]
     \centering
     \includegraphics[width=0.9\columnwidth]{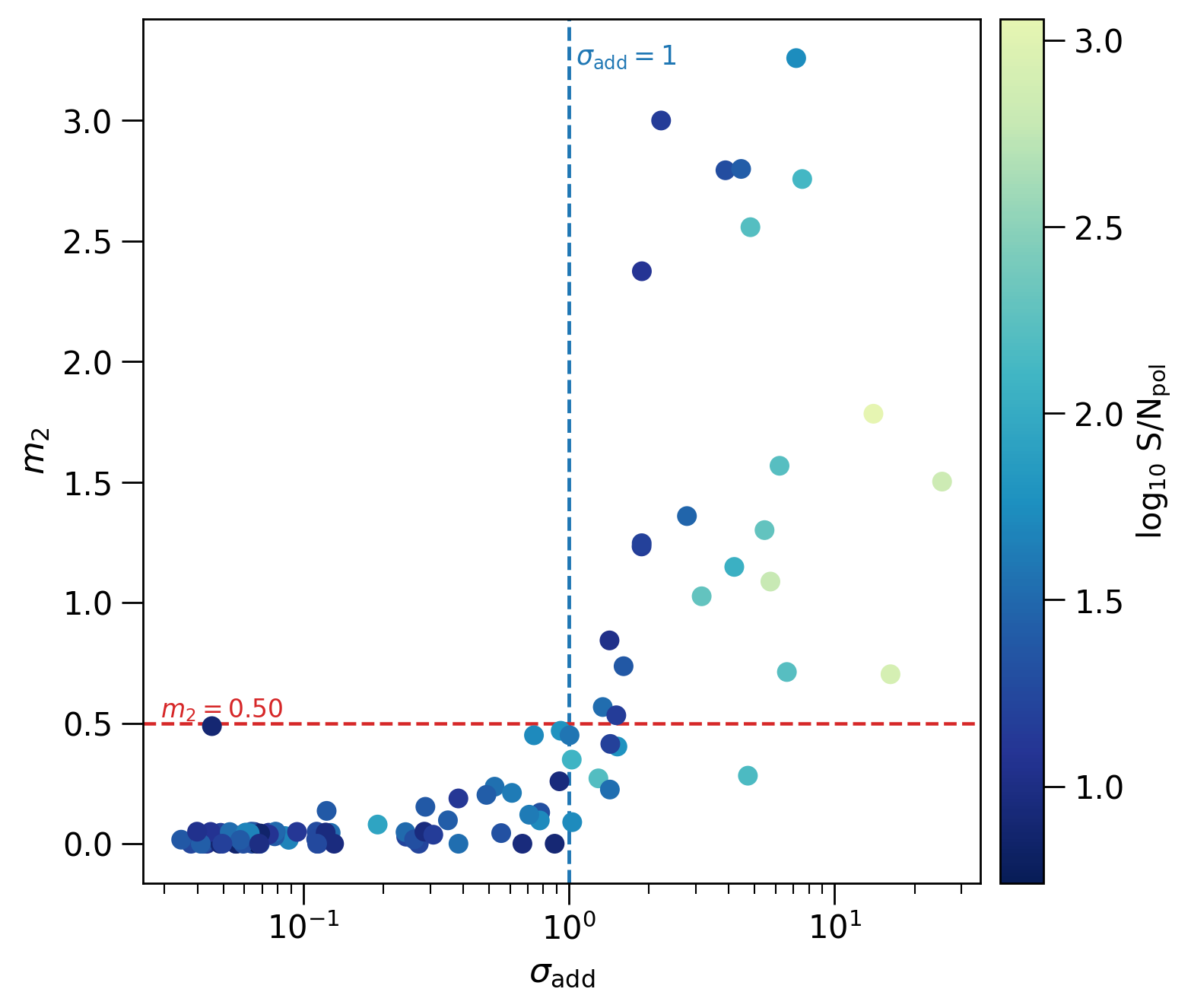}
     \caption{Comparison of the Faraday complexity metrics $m_2$ and $\sigma_\mathrm{add}$ for the sources in the sample to check for Faraday simplicity. The blue and red dotted lines show the limits below which sources are considered Faraday simple.}
    \label{fig:faraday_complexity}
\end{figure}

One of our outputs from the \texttt{ViMS} pipeline is an RM map across the field, assuming that all sources within it are Faraday-simple, i.e. the rotation of the polarisation vector is given by a single Faraday screen in front of the source. In this case the observed RM is a probe for the intervening medium (ICM, Milky Way). However, any significant contribution to the Faraday screen from the source itself or other intervening media can create more complex structures in Faraday depth, where the peak does not necessarily correlate with the RM of the ICM. To check whether the sources have a single Faraday peak, we employ the strategy of \cite{2024AJ....167..226V} and use two metrics to characterise the Faraday complexity. The first one being the normalised second moment of the clean peaks, $m_2$. This value is calculated from the clean components returned by \texttt{RMClean} and their FWHM in Faraday depth. The second metric we use is the complexity $\sigma_\mathrm{add}$ explained in \cite{2026ApJ...996...81V}, which measures the complexity in the QU spectra. Both metrics quantify how strongly the cleaned Faraday spectrum in the pixel of highest polarised intensity deviates from a single Gaussian. The higher the both values, the more complex is the Faraday spectrum of the observed source. For the threshold values of Faraday complexity we adopt the values of \cite{2024AJ....167..226V}, setting $m_2 \leq 0.5$ and  $\sigma_\mathrm{add} < 1$. With these thresholds we consider sources as complex as soon as more than one peak is resolved in Faraday depth. The result of this comparison is shown in Fig.~\ref{fig:faraday_complexity}.  
Based on this classification 75 out of the 101 sources are considered Faraday simple, while 26 do not fit that category. Therefore most of our sources (approximately 75\%) are Faraday simple at the observed wavelengths. This includes M\,49. 

\begin{figure}[ht]
     \centering
     \includegraphics[width=0.9\columnwidth]{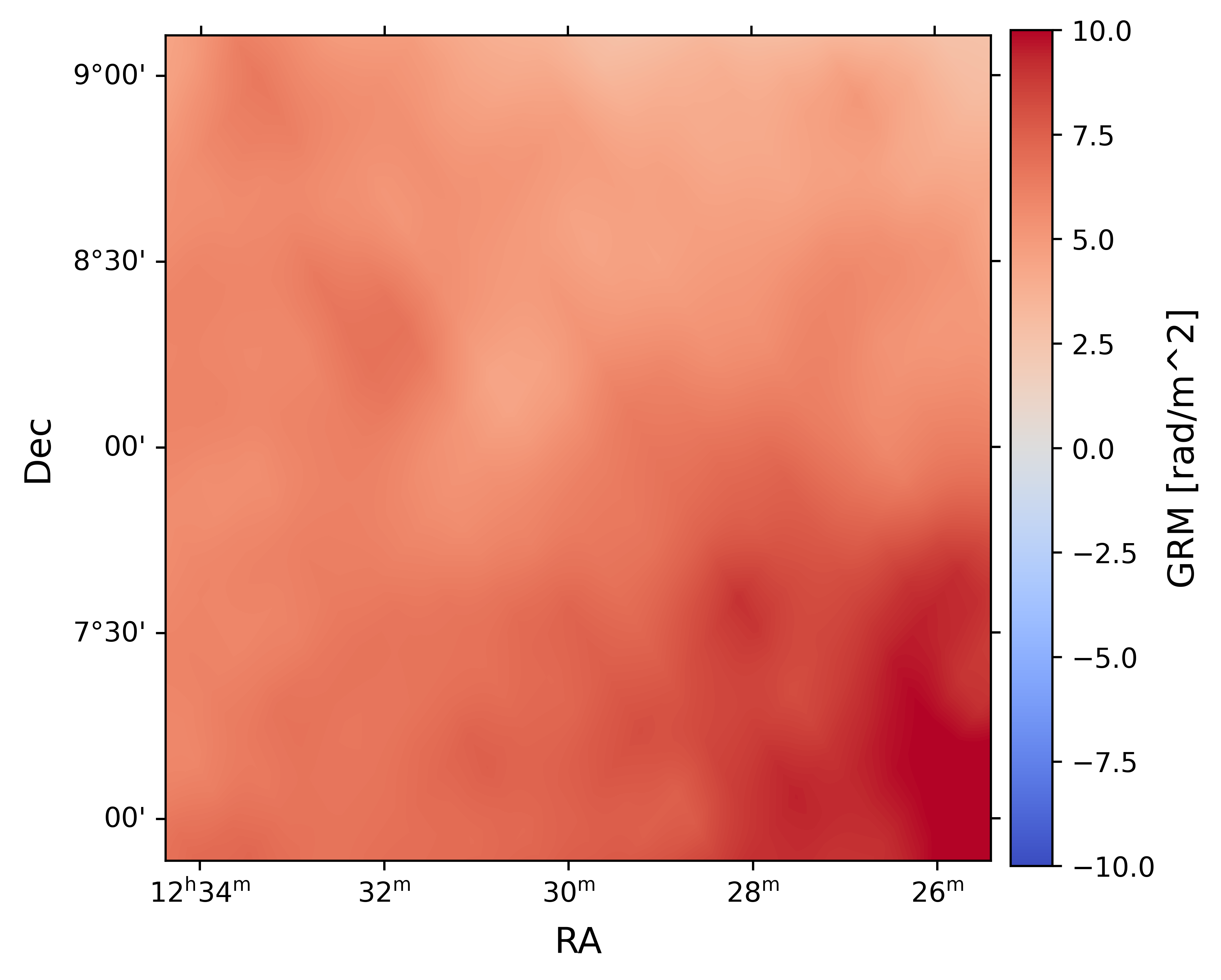}
     \caption{Galactic Rotation Measure map as derived from the 'Hutschenreuter map'. The median and standard deviation of the GRM are: $M = \SI{5.8}{rad/\meter^2}$ and $\sigma_{\mathrm{GRM}} = \SI{6.3}{rad/\meter^2}$.}
    \label{fig:GRM_Hutschenreuter}
\end{figure}

With this confirmation, we can assume that the RM values are caused primarily by the ICM of the Virgo cluster and by the Milky Way. To estimate the ICM contribution to the RM, we need to remove the effect of the Galactic RM (GRM). This is done using the foreground reconstruction by \cite{2022A&A...657A..43H}. Calculating the GRM contribution in the direction of M\,49, we determine the GRM map shown in Fig.~\ref{fig:GRM_Hutschenreuter}.
We determine a median value of $\SI{5.8}{rad/m^2}$ and a standard deviation $\sigma_{\mathrm{GRM}} = \SI{6.3}{rad/\meter^2}$ across the field using a radius of $\SI{1.5}{deg}$ for the GRM detection.   

\begin{figure}[ht]
     \centering
     \includegraphics[width=0.9\columnwidth]{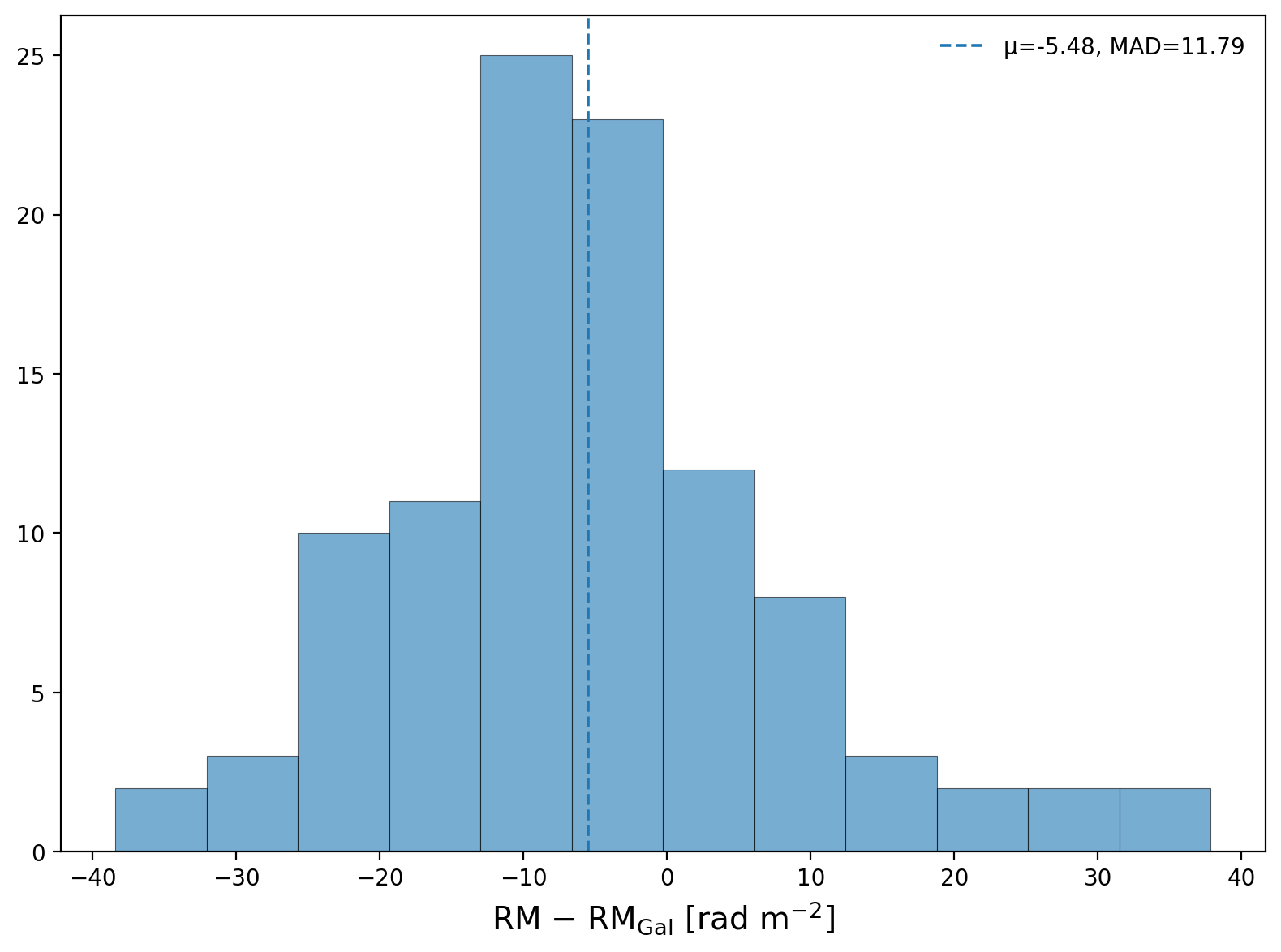}
     \caption{Residual RM distribution across the regarded field. The blue dotted line indicates the mean of the distribution ($\mu = \SI{-5.48}{rad/\meter^2}$). The RRM scatter is $\sigma_{\mathrm{RRM}} = \SI{11.79}{rad/\meter^2}$}
     \label{fig:RM_distr}
\end{figure}

After correcting for the Galactic RM by subtracting the thus created map, Fig.~\ref{fig:RM_distr} shows the residual RM (RRM) distribution for the field. For each detected source according to \texttt{SoFiA}, we calculate the polarised intensity-weighted mean of the RRM over all pixels within the source. This yields a single RRM value per source and its distribution shows that the RRM of the sources is clustered around 0 with variations between $-40$ and $\SI{+40}{rad/\meter^2}$. The mean value of the distribution is $\mu = \SI{-5.48}{rad/\meter^2}$. Using the Median Absolute Deviation (MAD), we calculate the scatter of the RRM to be $\sigma_{\mathrm{RRM}} = \SI{11.79}{rad/\meter^2}$.

\begin{figure}[ht]
     \centering
     \includegraphics[width=0.9\columnwidth]{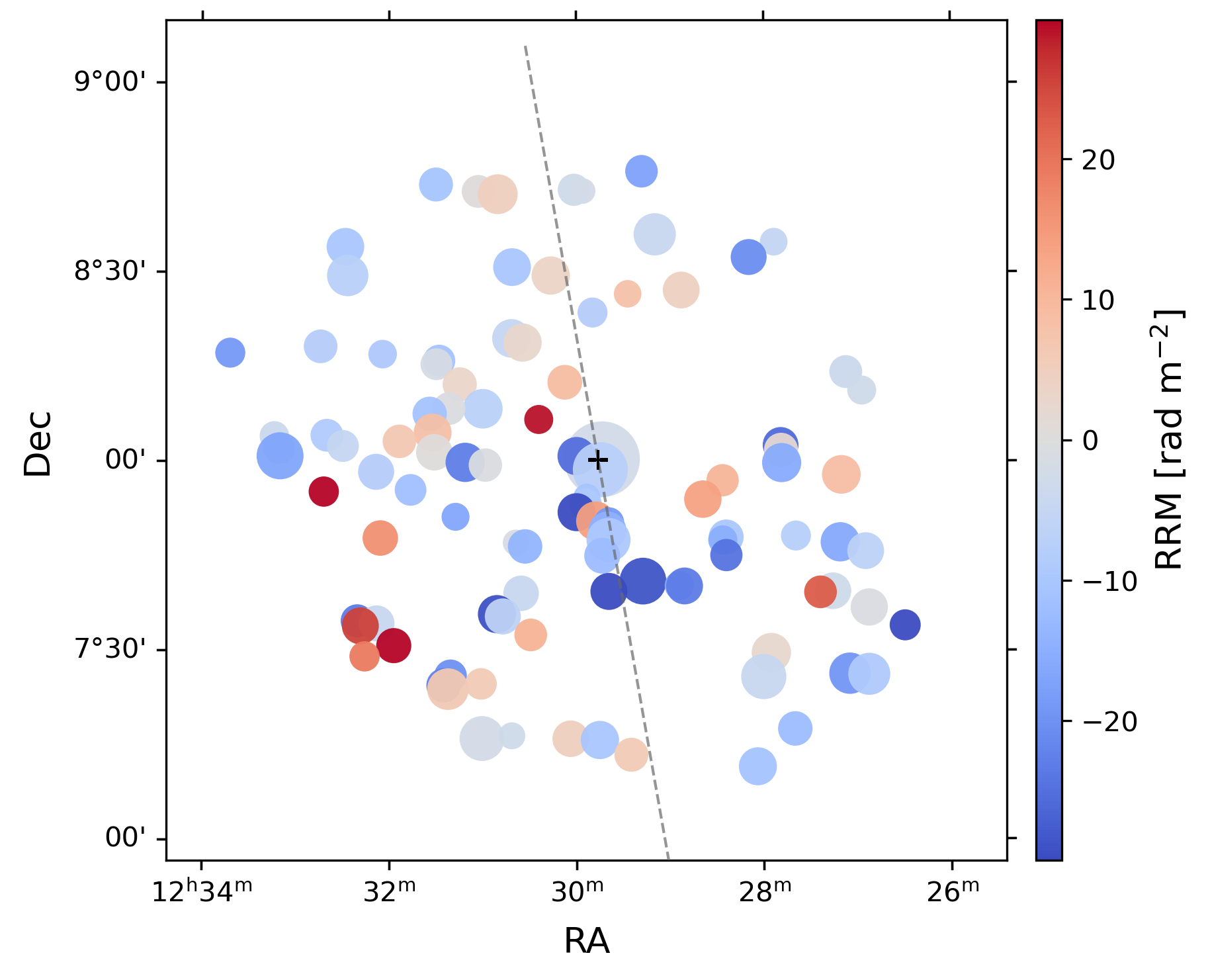}
     \caption{Residual RM map: The position of the points indicates their relative sky position and their radius represents the size on the sky according to the major axis output of \texttt{SoFiA} scaled by a factor of 30. The black cross shows the position of M\,49.The dotted line indicates the direction of the highest significance excess scatter.}
     \label{fig:RM_points}
\end{figure}

In Fig.~\ref{fig:RM_points}, we plot the sources in the field, with the colour scale showing the RRM for each individual source. Here we can see that the RRMs are distributed fairly uniformly across the sky without significant trends. Especially in the direction of infall of M\,49 (south-west towards north-east) there are no discernible differences among the RRM values. However, the scatter in RM seems to differ between the sources north and south of M\,49. To quantify this statement, we divide the field into two parts along an axis passing through M\,49 and determine the excess scatter between them. We vary this axis from 0° to 170° in steps of 10° going from North (=0°) through West (=90°), to determine if a direction of increased scatter exists. For this we need to determine an uncertainty of the scatter in each region, which we calculate following \cite{2025hsa..conf..108A}:
For each source we determine the uncertainty in RRM by propagating the error in the observed and the galactic RM. To estimate the scatter and uncertainty, we then employ a Monte-Carlo approach. In each of 10000 realisations we draw a simulated RM for each source from a Gaussian centred on the measured RRM of that source with a width equal to its uncertainty. For each realisation we determine the scatter using a robust standard deviation based on MAD. The final scatter and uncertainty are then taken as the median and MAD-based standard deviation of the scatter distribution across all realisations. We do this separately for each region around M\,49.
To determine whether the scatter between each pair of regions is significant, we follow \cite{2025hsa..conf..108A} and calculate the excess scatter for each Monte-Carlo realisation as:
\begin{equation}
    \sigma_{\mathrm{excess,}k} = \sqrt{\sigma_{\mathrm{south},k}^2 - \sigma_{\mathrm{north},k}^2}.
\end{equation}
From this we quantify the significance as the median excess scatter divided by its MAD-based standard deviation.

We determine the highest significance of excess scatter for two regions along the gray dotted line in Fig.~\ref{fig:RM_points}. The axis along which the field is split in two is perpendicular to it. The northern region contains 45 out of the 101 polarised sources and exhibits a median RRM of $\SI{-4.7}{rad/\meter^2}$. This is quite similar to the region towards the south where we find a median RRM of $\SI{-9.0}{rad/\meter^2}$ for 56 sources. In scatter the two regions show a larger difference with $\sigma_{\mathrm{RRM,North}} = \SI{8.7 \pm 2.1}{rad/\meter^2}$ in the north and $\sigma_{\mathrm{RRM,South}} = \SI{14.8 \pm 2.2}{rad/\meter^2}$ in the south. The excess of the RRM scatter south of M\,49 compared to the north is $\SI{11.9 \pm 3.1}{rad/\meter^2}$ which corresponds to a significance of $3.8\sigma$. As visually suspected, the scatter along a line close to the N-S direction seems to show a significant excess scatter. Similarly the direction along the north-south shows a scatter with a 3$\sigma$ significance, which matches the infall direction of M\,49 and could potentially be correlated with it. A plot showing the significance of the excess scatter along each direction and a table containing the calculated parameters for each region is shown in Appendix \ref{fig:dir_scatter}.

\section{Discussion}
\label{sec:discussion}

\subsection{ViMS pipeline performance and polarised source density}
In Sec.~\ref{sec:data_reduction} we introduced the ViMS pipeline used for calibration, imaging and RM synthesis of the data, focusing on the polarisation calibration. Producing reliable polarisation information is therefore of main interest. As shown by the calibration diagnostics in Appendix \ref{sec:pipeline_details}, the solutions are generally well behaved. When applied to the polarisation calibrator, the behaviour is as expected although some regions with leftover RFI are still visible, which indicates that the flagging strategy can still be improved. The derived Q, U and V values from the calibrator source are all within $3\sigma$ of the input model values. The measured polarisation fraction of the polarisation calibrator agrees well within 1$\sigma$ with the model by \cite{LinPolCal2024Memo} and does not vary by more than 0.1\%. The polarisation angle deviates by about $3.3 \sigma$ from the model value, but only within a fraction of a degree in between each other. The offset in polarisation angle in regards to the model might be connected to the uncertainty in the ionospheric RM calculation of $\sim \SI{0.4}{rad/m^2}$ which corresponds to 2 degrees at $\SI{1.4}{GHz}$. In that case, we would expect a distribution around the model value, but our polarisation angles are consistently smaller than the model's and only vary by 0.3 degree between the three different observations. This could be a coincidence given that only three observations are regarded. However, \cite{2025A&A...694A.125L} report an offset of approximately seven degrees in the polarisation angle, regardless of the used calibration tool (\texttt{CASA}, \citealp{2022PASP..134k4501C}; \texttt{CARACal}, \citealp{2020ASPC..527..635J}) which is not mitigated when the ionospheric RM is taken into account. We could be seeing a reduced version of this effect in our data. 

Relative to other MeerKAT calibration workflows, the ViMS pipeline follows similar cross-calibration steps. Its strength lies more in the treatment of the polarisation calibration, with multiple flagging steps specifically for the crosshand visibilities and its inclusion of ionospheric contributions. With the subtraction of M\,87, it also takes the local conditions of the Virgo cluster into account, during the calibration of each field. We have seen that this reduces the noise in total emission by $\sim \SI{2}{\mu Jy/beam}$ and therefore effectively increases the sensitivity in the field.  
The pipeline, however, follows a single strategy for calibration and requires a careful RFI mitigation before and during calibration and therefore manual inspection unlike a more automated calibration. It is also debatable whether some steps implemented in the pipeline improve the calibration, namely the setting of an accurate model for the polarisation calibrator. In our setup we corrupt the polarisation calibrator model with the ionosphere to solve the crosshand phase and delay against it. Comparing this to a calibration run without corrupting for the ionosphere reveals that the changes in the derived RM of the calibrator are negligibly small (the difference is approximately $\SI{0.1}{rad/m^2}$).

After applying the reduction strategy, we check each field for flux scaling issues, by calculating the flux ratio of sources also detected in RACS-low. Here, the ratios match with the expected values within $1\sigma$ as well, confirming that the calibration in amplitude is stable across different observations. After mosaicking the field we determine a mean noise in the Stokes $I$ mosaic of $\SI{12.6}{\mu Jy/beam}$ at a beam size of $13''$. Considering only the central region where pointings fully overlap, the noise drops to $\SI{7.4}{\mu Jy/beam}$. This is compatible with the expected noise level for these MeerKAT observations \citep{2025A&A...693A.189D}. The fields used in this analysis, however, do not make use of the full hexagonal structure of the pointings. Assuming seven fields, all with similar individual noise levels overlapping as given by the hexagonal grid will increase the sensitivity in the centre to approximately $\SI{5}{\mu Jy}$. The present mosaic therefore likely presents a conservative estimate of the final survey depth.  

For the final polarisation map we determine a mean Stokes $QU$ noise of $\SI{23.3}{\mu Jy/beam}$ with minimum value of $\SI{6.6}{\mu Jy/beam}$ in the central region, as well at a resolution of $13''$ for an observation time of 180 minutes on source (45 minute per pointing). At this sensitivity we detect a polarised source density of $\sim \SI{50}{sources/deg^2}$ in the centre, which decreases to $\SI{40}{sources/deg^2}$ towards the edge of the mosaic. These values are broadly consistent with other recent polarisation surveys. In the MeerKAT Fornax Survey, with each pointing receiving 10 hours on-sky, \cite{2025A&A...694A.125L} detected a source density of $\sim 80$ polarised sources per deg$^2$ and a mean Stokes $QU$ sensitivity of $\SI{2.6}{\mu Jy/beam}$ with a minimum of $\SI{1.4}{\mu Jy/beam}$. The deeper integration per pointing in the Fornax survey naturally explains their higher density. In an early science release the MeerKAT International GigaHertz Tiered Extragalactic Exploration (MIGHTEE) survey detects $\sim \SI{16}{sources/deg^2}$ for an 8-hour per pointing observation \citep{2024MNRAS.528.2511T}. The median sensitivity reached in Stokes $QU$ is $\sim \SI{28.2}{\mu Jy/beam}$. Another large polarisation survey using the ASKAP radio telescope is the Polarization Sky Survey of the Universe’s Magnetism (POSSUM). \cite{2021PASA...38...20A} reported in an early POSSUM study a polarised source density of $\SI{27}{sources/deg^2}$ at a sensitivity of $\SI{30}{\mu Jy/beam}$ in a 10 hours observations per pointing. The expected source density for the full project is 45 $- \SI{60}{sources/deg^2}$ at an expected sensitivity of $\SI{12}{\mu Jy/beam}$ for observations in the combined band \citep{2025PASA...42...91G}. Another observation with the ASKAP radio telescope is done in the context of the ASKAP-RACS survey, where \cite{2023PASA...40...40T} detect approximately $\SI{4}{sources/deg^2}$ with a mean Stokes $QU$ sensitivity of $\SI{80}{\mu Jy/PSF}$. 

For the full ViCTORIA polarisation project, \cite{2025A&A...693A.189D} predict 2000 polarised sources across the $\SI{112}{deg^2}$ field ($\SI{18}{sources/deg^2}$). Our results suggest that this estimate is conservative. Extrapolating to the full survey area, we expect twice that number. 

\subsection{Polarisation properties of M49}

\subsubsection{Morphology and polarisation fraction}
The polarised intensity map of M\,49 shown in Fig.~\ref{fig:M49_opt_pol_overlay} reveals a polarisation structure that differs markedly from the total intensity morphology. The polarised emission of M\,49 is strongest at the inner radio lobes, while the central region, where Stokes $I$ peaks, is almost entirely depolarised, most likely due to unordered magnetic fields on scales smaller than the beam size \citep{1988ARA&A..26...93S}. Otherwise, the patchy and structured polarised emission we see in the lobes of M\,49 is expected for complex magnetic fields within radio lobes as shown by \cite{2014MNRAS.443.1482H}.

This behaviour is traced more closely in the fractional polarisation map of M\,49. The extended radio lobes exhibit fractional polarisation, with mean values of $\sim (29 \pm 5)\%$, notably high compared to the typical range of $0.1\% - 30\%$, with a mean of $6.2\%$ reported for radiative-mode radio galaxies by \cite{2015ApJ...806...83O}. In two regions coinciding with the onset of the bending of the radio tails, the fractional polarisation reaches values of up to $\sim (70 \pm 6)\%$. This is close to the theoretical maximum of $p_{\rm max} = (1 -\alpha)/(5/3 - \alpha) \approx 72\%$ for optically thin synchrotron radiation in a perfectly uniform magnetic field, where the spectral index $\alpha$ is assumed to be $-0.7$ \citep{1970ranp.book.....P}. Such extreme values indicate a strong ordering of the magnetic fields in that region.

The trend of increasing fractional polarisation towards the edges of lobes is well established, both in simulations \citep[see][]{2011MNRAS.417..382H, 2025MNRAS.539.1668S} and observations \citep{2019A&A...622A.209A}. During the inflation of lobes, shear and compression align the magnetic field tangentially to the lobe boundary, which corresponds to a magnetic draping of the lobe or the ambient field. Simulations of resolved radio galaxies in cluster environments indicate that this effect can increase the fractional polarisation of sources above 30\% \citep{2014MNRAS.443.1482H}. In the case of M\,49, the high fractional polarisation at the bending points is likely affected by the interaction of the radio lobes with the surrounding medium.

\subsubsection{Magnetic field structure}
The B-field vectors, seen in Fig.~\ref{fig:M49_MF} reveal a structured and asymmetrical morphology across M\,49. In the inner regions of the lobes the field is parallel to the jets, consistent with compression and shear alignment along the jet of the field \citep{2014MNRAS.443.1482H, 2025MNRAS.539.1668S}.

In the outer parts of the lobes an asymmetry emerges in the orientation and behaviour of the B-vectors. In the western radio lobe the vector orientation follows the morphology of the radio emission closely. When the radio tail bends the vectors bend with it, staying parallel to the local emission edge. This is consistent with the simulations of \cite{2014MNRAS.443.1482H}, who find that the polarisation position angle transitions from perpendicular to parallel alignment with the jet at increasing distance from the core, reflecting a shift from toroidal to longitudinal field structures.

The eastern lobe, however, behaves differently. Rather than following the bending of the radio lobe, the B-vectors keep their orientation from the inner regions until they are completely orthogonal to the emission. \cite{2014MNRAS.443.1482H} note that the perpendicular-to-parallel transition in field alignment correlates with decreasing polarisation. The lack of alignment with the lobe could therefore point towards a different physical process dominating this region. One possibility could be that the interaction of M\,49 with the intergroup medium could have introduced or maintained a toroidal field component at large distances from the core.
Potentially supporting that hypothesis is a study by \cite{2024A&A...690A.195S}, which analyses M\,49 at $\SI{144}{MHz}$. At low frequencies, where more of the diffuse emission in the radio arms is revealed, the eastern arm shows a narrow, collimated radio structure, at larger distances from the centre. They attribute this to a confined magnetic field structure, which could potentially align with the assumption of toroidal magnetic fields suggested here.

This magnetic-field asymmetry is mirrored in the fractional polarisation, whose highest values are reached only in the western lobe. This is likely connected to the interaction of M\,49 with its environment and the different processes that govern each lobe. \cite{2026ApJ...996...81V} analyse the polarisation properties of radio galaxies with bent radio tails and how they are influenced by the bending angle. They find a correlation between the B-vectors and the radio morphology, but no correlation between the RM and the bending angle. Our results for M\,49 are consistent with both findings. We see the correlation between the B-field and morphology, caused by the galaxies' motion through the surrounding medium and see no effect of the bending on the Faraday rotation.

\subsubsection{Rotation measure across M\,49}
In the RM map of M\,49, no significant variations across the lobes can be seen, despite the asymmetries in other polarisation properties. One would expect an RM enhancement at lobe edges in regions with higher polarisation fraction, due to the compression of the surrounding medium by the infall of the galaxy. This has been predicted in simulations of radio galaxies in cluster environments \citep{2011MNRAS.417..382H}. We do not observe any enhancement towards the rims of the lobes. The mean RM in the region where M\,49 resides is very low. Although RM depends on both electron density and line-of-sight magnetic field strength, these small values indicate weak magnetic fields in this part of the cluster. An increase in RM towards the rim of the lobes, even if it is as high as 70\% as estimated by \cite{2011MNRAS.417..382H}, would only be around 5 -$\SI{7}{rad/m^2}$. This is on the scale of scatter seen within each lobe (6 -$\SI{7}{rad/m^2}$), and is therefore not detectable for us. 

The absence of RM variations while, at the same time, observing asymmetries for the two lobes in polarisation properties indicates that local Faraday screens of M\,49's immediate environment contribute little to the observed RM. It seems that the dominant Faraday-rotating medium is the ICM of the Virgo cluster, which shows small variations across the extent of M\,49. This is consistent with the result of \cite{2004JKAS...37..337C}, that the dominant factor contributing to the RM of sources is the ICM with local Faraday screens being of less importance. It also agrees with the analysis by \cite{2026ApJ...996...81V}, who find no correlation between the RM of bent-tail radio galaxies and their bending angle. 

\subsection{The RM distribution in the surrounding region}
The M\,49 group lies in the outskirts of the Virgo cluster where electron density and magnetic field strength are less pronounced \citep[$n_e \approx \SI{5e-4}{cm^{-3}}$, ][]{2019AJ....158....6S}. Hence, we would expect a low RM of O(10 rad/m$^2$) and a low scatter of the RM. \cite{2001ApJ...547L.111C} and \cite{2010PhDT.......259P} found their values to be on the order of $\langle \mathrm{RRM} \rangle = \SI{20}{rad/\meter^2}$ with $\sigma_{\mathrm{RM}} = \SI{10}{rad/\meter^2}$. Our scatter is similarly low, while our mean RRM of $\SI{-5.48}{rad/m^2}$ is suspiciously close to 0, especially since instrumental leakage would appear at a Faraday depth of exactly zero. We have verified the robustness of the results by testing the Faraday simplicity of the sources using two independent Faraday complexity metrics, confirming that after cleaning, approximately 75\% of the background sources show a single peak in Faraday depth, without any secondary peaks of comparable amplitude.

A physical explanation of the low values is that in a turbulent magneto-ionic medium like the ICM, the magnetic field varies along the line of sight, causing different RM contributions (positive and negative) to overlap and partially cancel out \citep{2010A&A...522A.105G}. The mean RM therefore tends towards zero while the scatter indicates the turbulence of the medium. Hence, the low mean RM and small scatter are consistent with a weak turbulent magnetic field. 

\subsection{Interaction between the M\,49 group and the ICM}
The asymmetry in the RM scatter between the region north and south of M\,49 has a significance of $3.8\sigma$. The other highest significances are found in the directions right next to it. The directions towards the south-west show a significance slightly below $3\sigma$, while no excess-scatter is seen in the northern directions. This falls in line with the observed infall direction of M\,49, from south-west towards north-east. The region for which we find the highest excess-scatter corresponds to the wake behind the infalling galaxy and group. The enhanced scatter in the south could be connected to this infall and might trace the disturbance of the magnetic field in the ICM. \cite{2024A&A...689A.113M} have noticed in their X-ray map an emission halo more extended towards M\,49's southern edge compared to its northern one, indicating an asymmetric gas distribution, due to stripping (see also \cite{2019AJ....158....6S}). Simulations of galaxies falling into the clusters predict the formation of a turbulent wake behind the infalling object \citep{2017ApJ...848...27K}. Sloshing cold fronts, such as the one observed in M\,49 \citep{2019AJ....158....6S}, are associated with large-scale gas motions that can further distort and randomise the magnetic field orientation \citep{2004A&A...426..387S}. \cite{2025arXiv251026218L} simulate the influence of sloshing motions on magnetic fields and RM observables, finding that gas motions behind sloshing fronts produce small-scale RM fluctuations that increase the scatter among background sources. This would be consistent with our observations.

A further environmental effect could arise from the interaction of M\,49 with the dwarf galaxy VCC\,1249, located south-east of it. A stripped tail is seen in near ultraviolet and in HI maps, extending towards M\,49 \citep[Lee et al., in prep.]{2012A&A...543A.112A}. As the RM excess-scatter is slightly more pronounced in the south-west direction, we do not identify it as the main driver of the excess scatter. Nevertheless, it may contribute to the turbulence in the southern region.

\cite{2019AJ....158....6S} determined a virial radius for the group of 740\,kpc, which is larger than the region we analyse here. They associate the interactions they find in the X-ray around M\,49 to the movement of the galaxy against the group gas. An interesting question is therefore whether the enhanced scatter in RRM traces the turbulence in the group of M\,49 or in the ICM of the cluster.
An important constraint to that question is given by the analysis of the RM distribution across M\,49 itself. This distribution is very uniform across the lobes, without any correlation to its morphological asymmetries. Since the radio lobes of M\,49 are embedded in the intragroup medium, we would expect to see any variations in RM across the lobes if it would trace the groups variability. We can therefore assume that the group medium is not a significant Faraday screen, and the contributions to the RM are influenced mainly by the ICM. The scatter in RM in the wake of M\,49 is thus tracing the disturbance in the ICM due to the passage of the group.

It appears that during its infall the M\,49 group has interacted with the ICM and created a turbulent region in its wake, which produced the higher RM scatter. The group medium itself does not seem to contribute to the Faraday rotation in a significant way. However, we cannot fully exclude the possibility of the scatter being connected to the group instead of the ICM. Mapping a larger region will let us compare the RRM scatter within the virial radius of the group and outside of it. The extension of this study to all of the Virgo cluster (Spasic et al., in prep.) will create a clearer picture. 

\section{Conclusions}
\label{sec:conclusions}

In this paper we have analysed a pilot field around the infalling galaxy M\,49 and its group in the context of the ViCTORIA project. The main conclusions are as follows:

\begin{enumerate}

\item Using MeerKAT L-band full-polarisation data calibrated with the ViCTORIA MeerKAT Survey (ViMS) pipeline, we obtained a deep polarimetric view on the M\,49 region of the Virgo cluster. With four pointings covering $\sim 2.5\,\mathrm{deg}^2$ around M\,49, we reach a noise level of $23.3\,\mu\mathrm{Jy/beam}$ in polarisation at a resolution of $13''$. We detect 101 polarised background sources, yielding a source density of $\sim 50\,\mathrm{sources/deg}^2$ in the inner mosaicked region and $\sim 40\,\mathrm{sources/deg}^2$ in the outer regions. This yields an RM-grid resolving scales in the kpc regime. Extrapolating to the full $112\,\mathrm{deg}^2$ ViCTORIA survey, this implies more than 4000 polarised sources, exceeding the original survey predictions by a factor of two.

\item The galaxy M\,49 shows polarised emission that is extended and highly structured, exhibiting a morphology markedly different from the total intensity. The elevated fractional polarisation at the bending points is consistent with compression and ordering of the magnetic field due to the effects of ram-pressure exerted by the ICM. In the western lobe, the B-vectors follow the bending of the radio morphology. In the eastern lobe, the B-vector orientation is toroidal well into the extended tail region, possibly due to confinement by the higher ambient density.

\item The RM across the lobes of M\,49 is uniformly low, with values between $-25$ and $+25\,\mathrm{rad\,m}^{-2}$ and no significant difference between the eastern and western lobe. The residual RM scatter is higher in the southern region with a significance of $3.8\sigma$, potentially caused by the turbulent wake behind M\,49. This demonstrates the power of RM analysis to study the non-thermal properties of the ICM.

\end{enumerate}

The forthcoming full ViCTORIA polarisation survey of the Virgo cluster (Spasic et al., in prep.) will extend this analysis to the full $112\,\mathrm{deg}^2$ field, enabling a more detailed mapping of the magnetic field structure across the cluster.

\begin{acknowledgements}
AS and MB acknowledge support by the Deutsche Forschungsgemeinschaft (DFG, German Research Foundation) under Germany’s Excellence Strategy – EXC 2121 „Quantum Universe“ – 390833306. MB also acknowledges the DFG Research Group "Relativistic Jets" FOR5195 – project number 443220636.
FdG and AB acknowledges support from the ERC Consolidator Grant ULU 101086378. VG acknowledges support by the German Federal Ministry of Education and Research (BMBF) under grant D-MeerKAT III.
The MeerKAT telescope is operated by the South African Radio Astronomy Observatory, which is a facility of the National Research Foundation, an agency of the Department of Science and Innovation.
\end{acknowledgements}

\bibliographystyle{aa}
\bibliography{bibfile}

\appendix
\begin{appendix}

\section{Details of the ViMS pipeline}

\begin{figure*}
   \resizebox{\hsize}{!}
            {\includegraphics[scale=0.8]{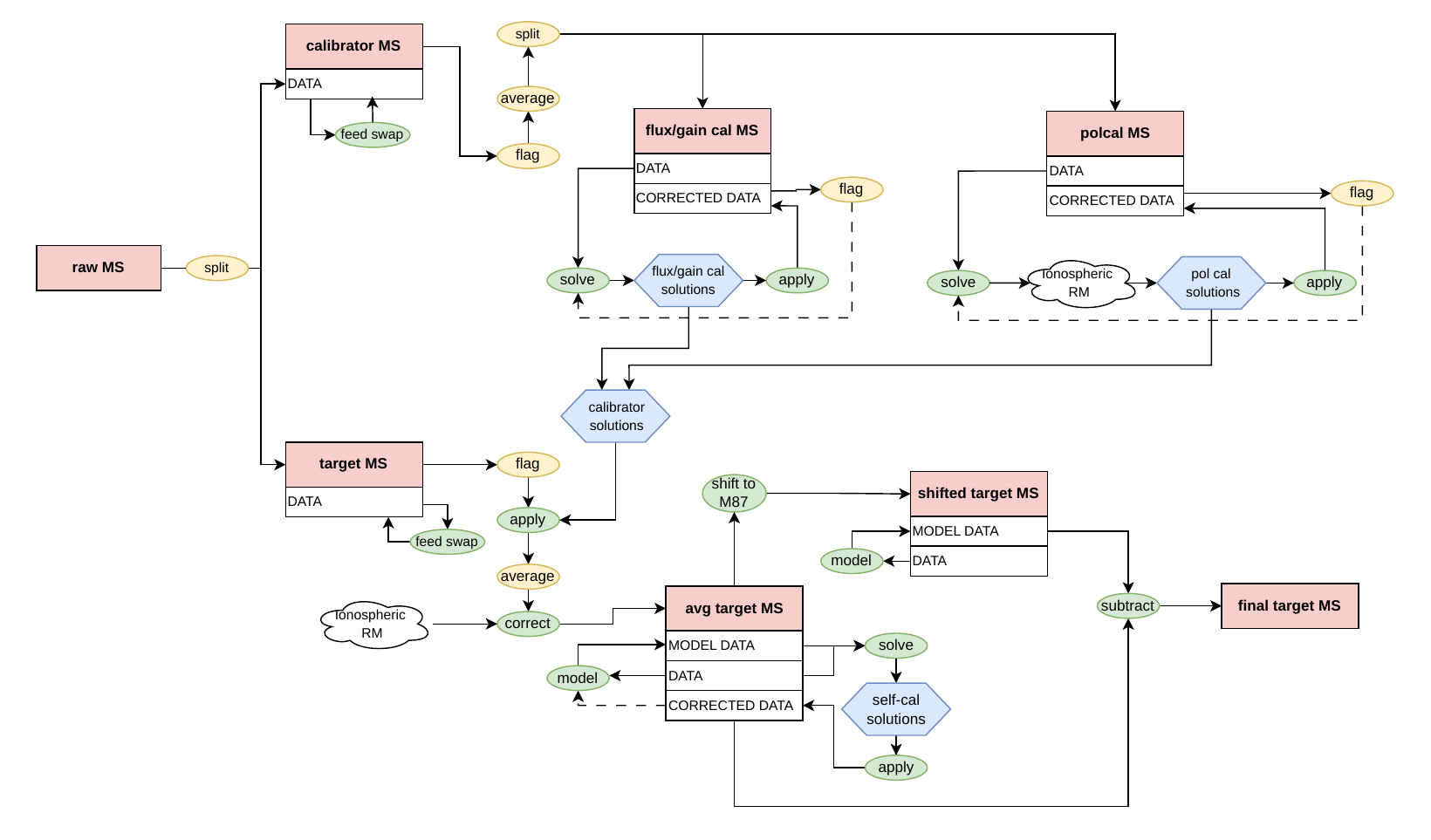}}
    \caption{Schematic of the calibration process in the ViMS pipeline as used to calibrate the pilot fields.}
    \label{fig:vims_overview}
\end{figure*}

\label{sec:pipeline_details}
In this section we explain the details of the ViMS pipeline and how the processing packages are combined. The ViMS pipeline is a Python-based workflow, which calls a sequence of software packages during execution as command-line processes instead of re-implementing their algorithms. The main packages used this way are \texttt{AOFlagger} \citep{2010ascl.soft10017O} and \texttt{tricolour} \citep{2022ASPC..532..541H} for flagging, \texttt{DP3} \citep{2018ascl.soft04003V} and \texttt{Spinifex} \citep{mevius_2025_15000430} for data averaging and ionospheric Faraday-rotation corrections, \texttt{facetselfcal} \citep{2021A&A...651A.115V} for self-calibration, \texttt{WSClean} \citep{2014MNRAS.444..606O} for imaging, \texttt{MosaicQueen} \footnote{\url{https://github.com/caracal-pipeline/MosaicQueen}} for mosaicking and \texttt{RM-Tools} \citep{2026arXiv260120092V} for RM synthesis. These packages are executed as command-line processes. An exception to this is \texttt{CASA} \citep{2022PASP..134k4501C}, used for calibration steps and preparation, whose tasks are called directly from Python. Each step is executed inside a custom singularity in order to exploit different packages and software \footnote{\url{https://hub.docker.com/repository/docker/abenati/vims}}.

The pipeline can be started via Python command, which will then loop through each ViCTORIA MeerKAT observation and create an output directory structure per observation. The pipeline can also be started and stopped on each processing step and run for individual observations, to help with debugging or testing without running the full workflow. For each processing step of the pipeline intermediate data products are written to disk, sorted by their use in the directory structure. These include corrected measurement sets, calibration tables, diagnostic plots and image products. The ViMS pipeline therefore handles the formatting of the data products to be used in each of the software commands.

In addition the ViMS pipeline creates a log file for each observation, tracking the exact steps and the output of the software and stores diagnostic products to assess the quality of calibration and imaging. These include a tracking of succeeded or failed steps, flagging percentages and image statistics. In the following sections the steps and parameter settings of the pipeline are explained in more detail. In Fig.~\ref{fig:vims_overview} an overview of the pipelines calibration steps is shown.

\subsection{Data preparation and flagging strategy}
The starting point of the reduction process is the raw MeerKAT measurement sets. The ViMS pipeline splits the data into separate target files, a measurement set containing the flux and gain calibrator and a separate file containing the polarisation calibrator using the \texttt{CASA} task \texttt{mstransform}. As \cite{EVLA2022Memo} have shown, the handedness of the MeerKAT data shows a discrepancy with the IAU definition, which we correct for using a specialised script \footnote{\url{https://github.com/bennahugo/LunaticPolarimetry/blob/master/correct_parang.py}}. With this an anti-diagonal matrix is applied to the raw visibilities. Furthermore we set the receptor angle (i.e. the reference angle for polarisation) to zero in the metadata. This is done for each observation and logged to prevent any accidental multiple conversions. Additionally, the low and high frequency edges are discarded by limiting the frequency band to above \SI{900}{MHz} and below \SI{1.65}{GHz}, to avoid band-pass roll-offs.

For all measurement sets, we first flag the autocorrelations, the shadowed antennas and the frequency range of the HI line of the Milky Way ($1419.8 - \SI{1421.3}{MHz}$) in the data. Additionally, the antenna `m041' is removed from the data in the first program (the 2022/23 cycle), as it shows diverging solutions and an odd behaviour after calibration across multiple observations. This initial flagging of the data is done using the CASA task \texttt{flagdata}.

\begin{figure*}[h]
     \centering
     \begin{subfigure}{\columnwidth}
     \includegraphics[width=\columnwidth]{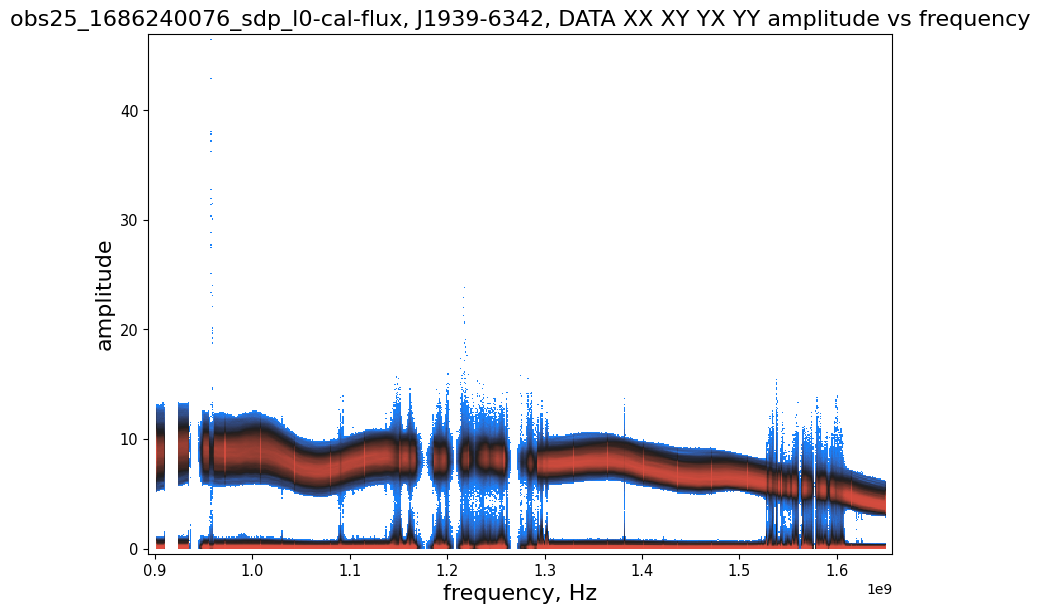}
     \caption{After initial flagging strategy.}
     \label{fig:fluxcal_flag}
     \end{subfigure}
     \begin{subfigure}{\columnwidth}
     \includegraphics[width=\columnwidth]{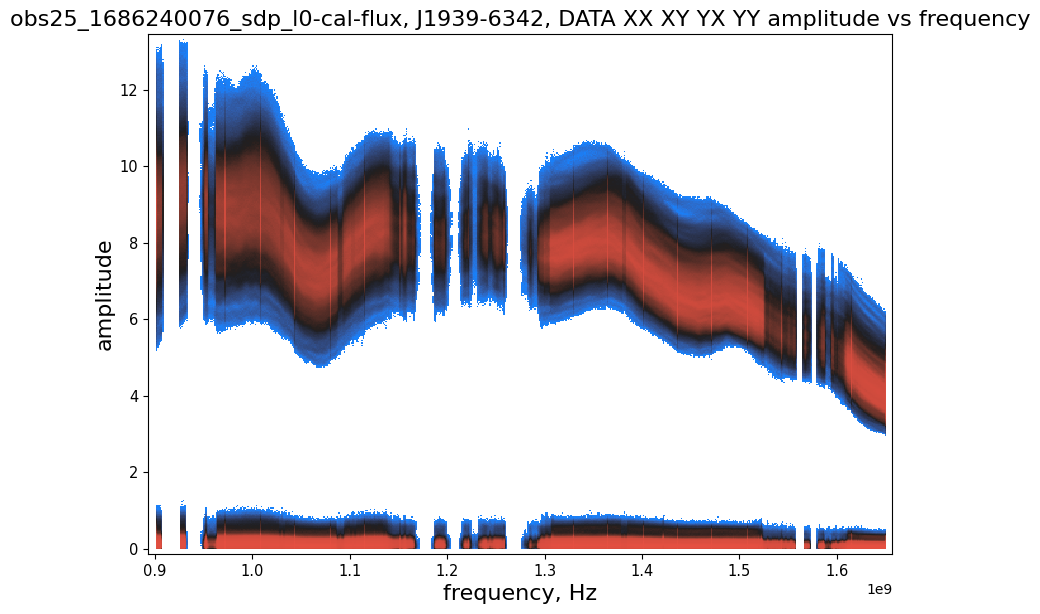}  
     \caption{After cross-calibration flagging.}
     \label{fig:fluxcal_flag_crosscal}
     \end{subfigure}
     \caption{Flux calibrator J$1939-6342$ after flagging.}
\end{figure*}

For the automated RFI excision we use different strategies for the calibrators and the target fields. For the calibrators we use \texttt{AOFlagger} to remove data affected by RFI \citep{2010ascl.soft10017O}, which we run twice, to ensure that most RFI is removed from the data. Starting from the default flagging strategy as given by \texttt{AOFlagger} \footnote{\url{https://gitlab.com/aroffringa/aoflagger/-/blob/master/data/strategies/generic-default.lua?ref_type=heads}}, we flag the complex values in amplitude for all products independently (HH, VV, HV, VH) and set the \texttt{base$\_$threshold} = 3. The other settings we leave as specified in the strategy. Employing this strategy we are able to remove most of the strong RFI regions in the data, with a flagging of approximately 30\% as can be seen example-wise for the flux calibrator in Fig.~\ref{fig:fluxcal_flag}. As can be seen some regions of RFI are still present in the data, however, the calibrator sources will be further flagged in between calibration steps. It is therefore important to not employ a too aggressive flagging strategy in the beginning to avoid overflagging during the calibration. 

If we include the RFI excision done during calibration the flagged percentage of the data is approximately 50\%. Afterwards, most of the RFI is removed as can be seen in Figure ~\ref{fig:fluxcal_flag_crosscal}.

\begin{figure}[h]
     \centering
     \includegraphics[width=\columnwidth]{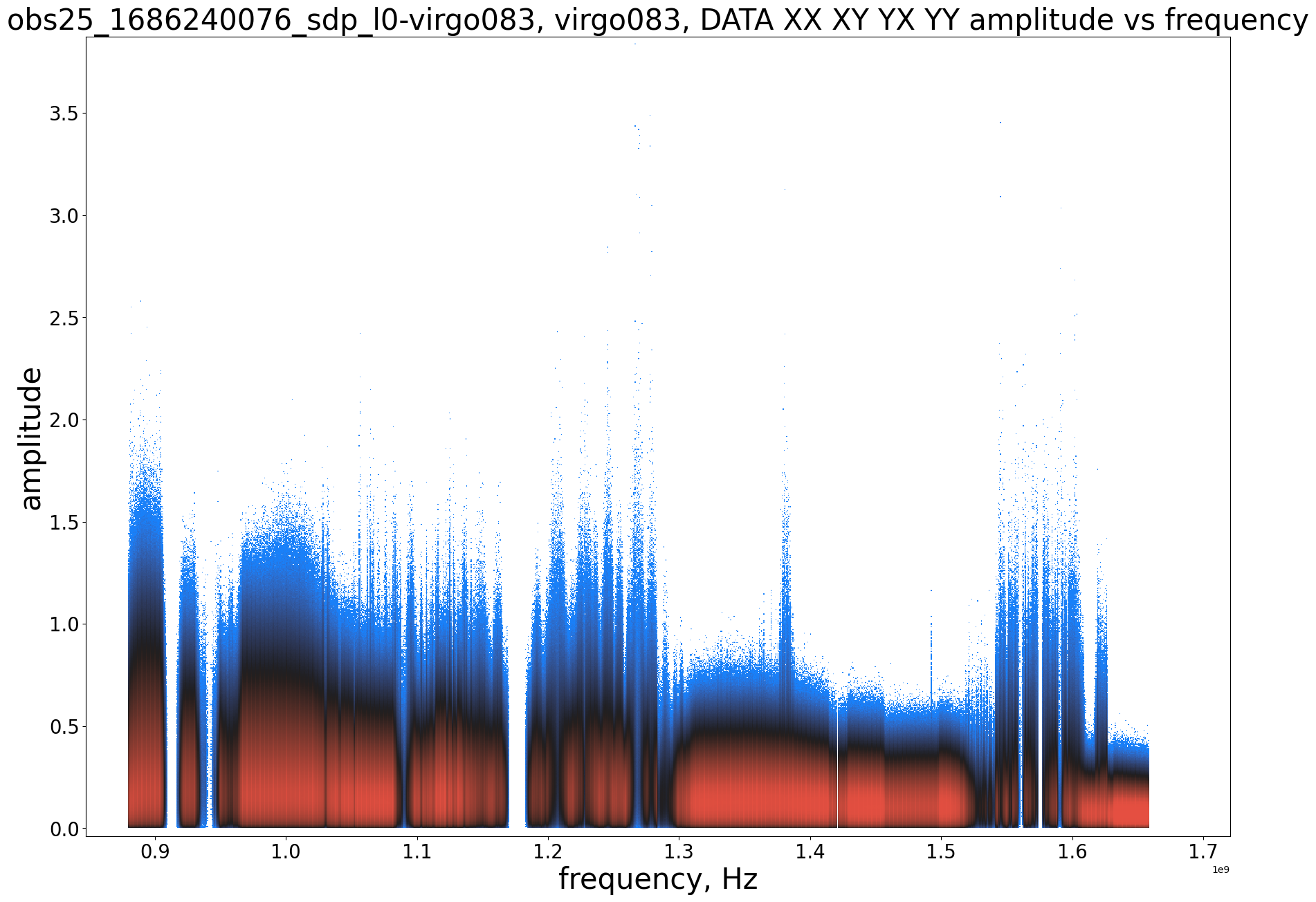}
     \caption{Target field after application of stricter flagging strategy with \texttt{tricolour}.}
    \label{fig:target_flag}
\end{figure}

For the automated flagging strategy of the target fields, we consider a slightly different approach. Instead of \texttt{AOFlagger} we use \texttt{tricolour} with a MeerKAT-specific flagging strategy for target fields taken from the \texttt{VermeerKAT} pipeline \footnote{\url{https://github.com/bennahugo/VermeerKAT/blob/c012dc016a7475e9bc7e09aa7ae6fd4abe2083cc/vermeerkat/data/input/mk_rfi_flagging_target_fields_firstpass.yaml}}. With this strategy the RFI excision is much stricter with about 50\% of the data in the targets flagged (see Fig.~\ref{fig:target_flag}). This is needed in case of the target fields, as they are not flagged in any further steps of the calibration process. The flagging rate of 50\% corresponds to the flagging rate of the calibrator fields after the additional flagging done during calibration.

After flagging the calibrator and target field measurement sets are averaged to 598 channels, corresponding to a channel width of $\SI{1254}{\kilo \Hz}$.

\subsection{calibration sequence}

As a first step to the calibration strategy we set a model for the bandpass calibrator J$1939-6342$ using the \texttt{CASA} task \texttt{setjy}, which sets the flux density for a point-source at the phase centre. With this task the visibility amplitude and phase associated with flux density scale are written to the model column where we are using `Stevens-Reynolds 2016' as the flux density standard. For L-band observation, sources close to J$1939-6342$ are relatively dim, therefore setting a simple point source model is sufficient for the reduction. We similarly set the model for the polarisation calibrator as a point source in the phase-centre using \texttt{setjy}. Instead of using a flux scale standard we set the model manually, implementing the model for the polarisation angle and polarised intensity of \cite{LinPolCal2024Memo}.

\begin{figure*}[h]
     \centering
     \includegraphics[scale=0.4]{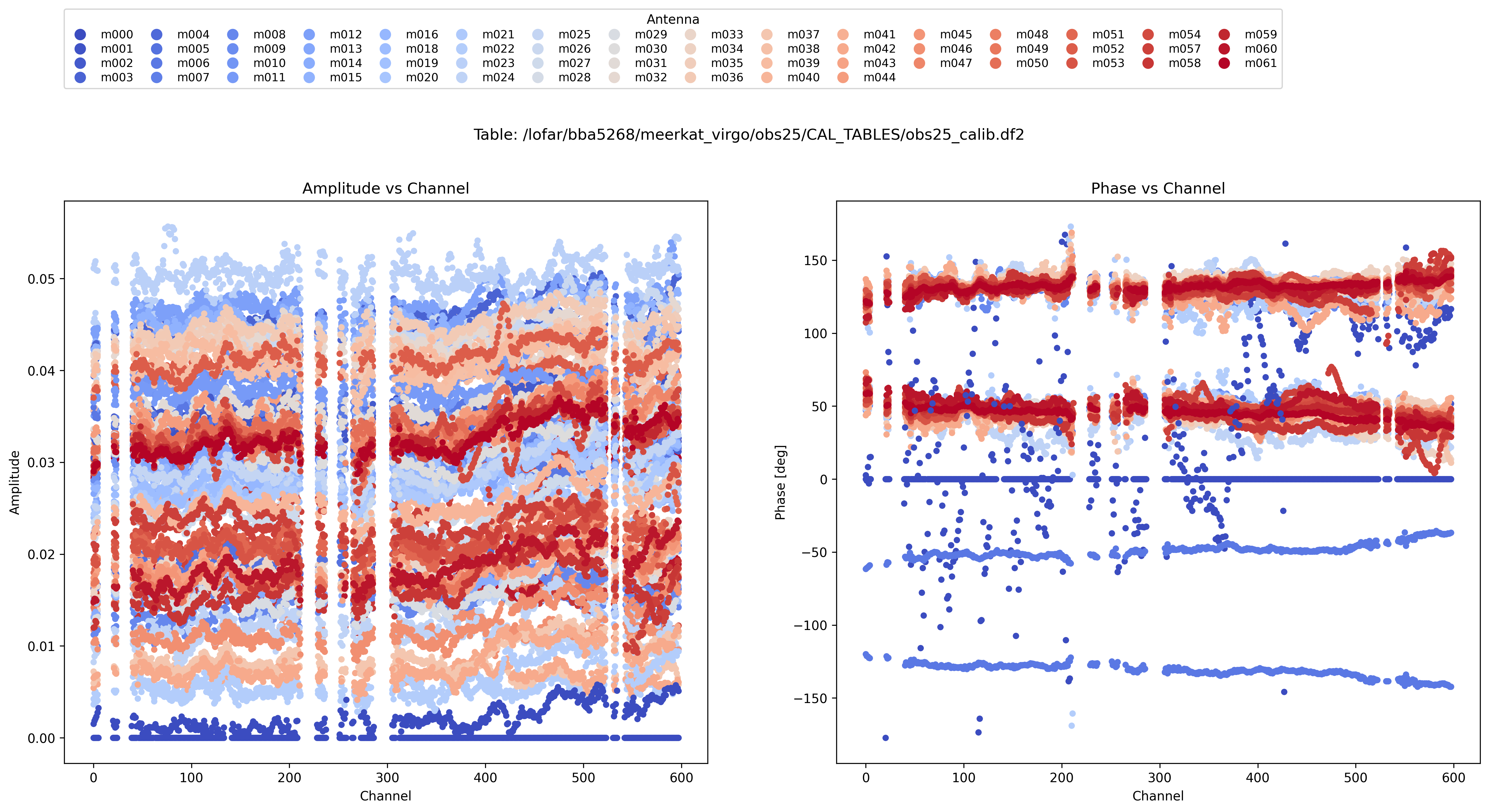}
     \caption{Leakage calibration solutions for a single observation. The left plot shows the amplitude solution, the right one the phase solution.}
    \label{fig:leakage_sol}
\end{figure*}

We perform the calibration of the parallel hands (HH, VV) in two rounds. In each iteration, we calibrate for the delay (K), the gains (where we separate the calibration for phase and amplitude, G) and the bandpass (B). After this first round we apply the solution tables of the first round to the bandpass calibrator and refine the RFI excision using \texttt{CASA}'s \texttt{tfcrop} on the corrected data column. This way we are able to remove RFI which becomes identifiable after most instrumental responses are removed. Then we repeat the same calibrations for the bandpass calibrator, to redetermine the solutions from the data. Similarly to the previous calibration, we perform the leakage calibration in two steps. We use the unpolarised bandpass calibrator J$1939-6342$ to calibrate the off-axis leakage using the linear least-squares method (Dflls). The leakage solutions are strongly affected by residual RFI in the data, specifically in the crosshands (HV, VH). We therefore flag the crosshands of the corrected bandpass calibrator data explicitly using the \texttt{CASA} task \texttt{tfcrop}, before deriving the first iteration of leakage solutions and also after applying them to the bandpass calibrator, before redetermining the solutions. We also flag the solutions of the leakage calibration at the edges of flagged channels, to avoid any boundary effects occurring. An example of the leakage solutions we derive is shown in Fig.~\ref{fig:leakage_sol}. The amplitude is expected to be small, and the phase should be approximately stable across frequency, both of which can be seen for our solutions. 
 
We start the calibration for the gain calibrator J$1150-0023$ by recomputing the delays for it. This is done by setting a solution interval of 30 minutes, to track potential changes across the 5 scans of the gain calibrator, which might remain in the data when the delay solutions of the bandpass calibrator (for which a single solution in time is determined) are transferred. For the gain calibrator the focus is on the phase calibration as these solutions will be transferred onto the target fields and therefore need to be as accurate as possible. Similar to the delay calibration we employ a two-step calibration of the phase including a self-calibration step. This is done by first determining a single phase solution across time (by setting \texttt{solint}='inf'), as a first stable approximation. We then apply this solution together with the previously determined ones to the gain calibrator, create a model of the gain calibrator using \texttt{CASA}'s \texttt{tclean} and recalculate a phase solution based on this model, using a shorter solution interval of 30 seconds. We then apply these newly created solutions to the gain calibrator, flag the visibilities based on the residual between the corrected data and the model, and redetermine the phase solutions afterwards for an again shorter solution interval of 15 seconds. This way we try to include potential temporal variabilities in the data, while still mitigating the effect of RFI. As we do not have an external model for the gain calibrator, we do not calibrate its amplitude but apply the amplitude solution of the bandpass calibrator and only determine its variations by determining scalar amplitude solutions (T) normalised to one. These solutions are together with the amplitude solutions of the bandpass calibrator applied to target fields in the end.

\begin{figure}[h]
     \centering
     \includegraphics[width=0.8\columnwidth]{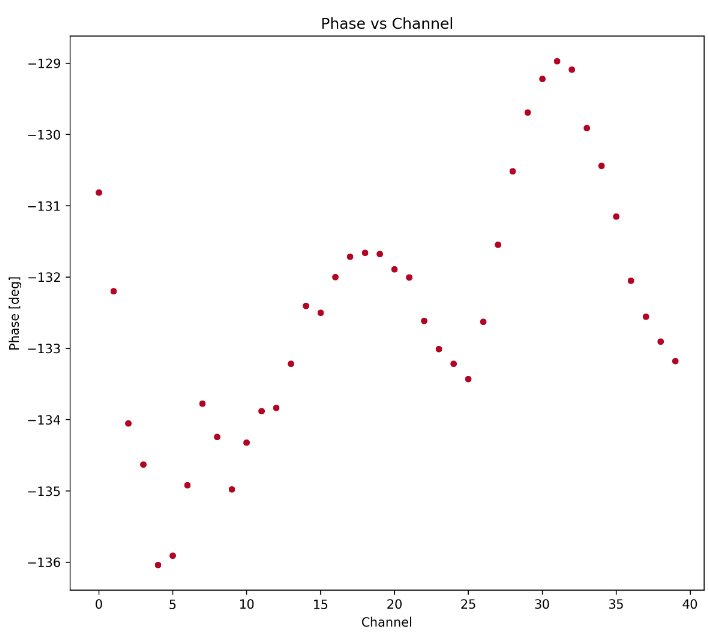}
     \caption{Crosshand phase solution for a single observation.}
    \label{fig:crosshand_phase}
\end{figure}

\begin{figure}[h]
     \centering
     \includegraphics[width=\columnwidth]{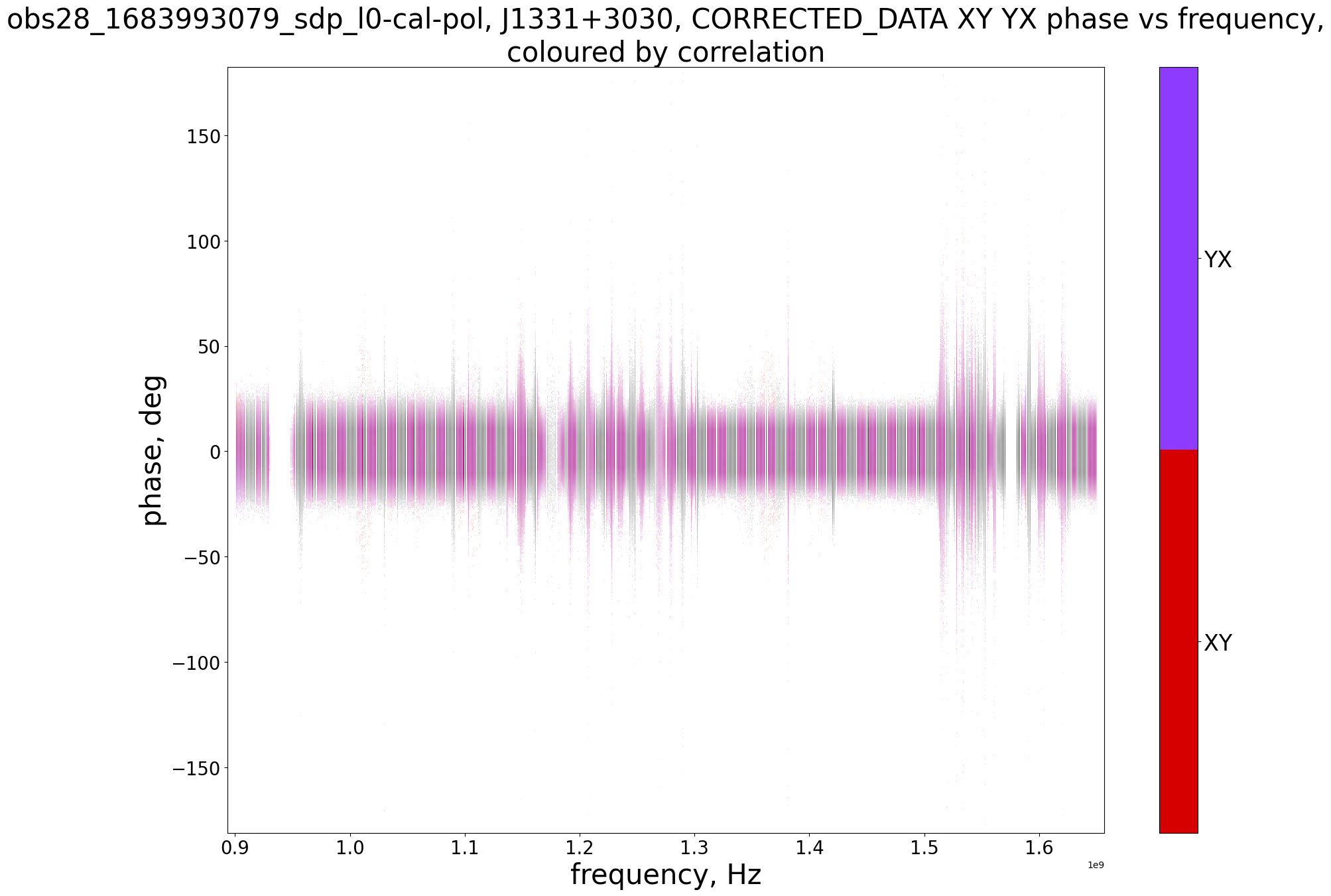}
     \caption{Phase frequency plot of the crosshand visibilities of the polarisation calibrator after application of all calibration solutions.}
    \label{fig:vis_phase}
\end{figure}

For the polarisation calibration we have set a frequency dependent full polarisation model. \cite{LinPolCal2024Memo} have derived a new model for the polarisation angle and polarised intensity of the polarisation calibrator 3C286 at frequencies below $\SI{12}{\giga\Hz}$, which we implement for the frequency range of the L-band. Computing the ionosphere with \texttt{spinifex}, we can corrupt the model for the effects of the ionosphere using \texttt{DP3}. The subsequent polarisation solutions are then derived against a model that already includes the expected ionospheric Faraday rotation. Following again the two step approach as before, we calibrate again for the delay and phase with solint = `inf', deriving a single solution in time. We then apply the solution together with the previous ones to the polarisation calibrator before flagging the visibilities and recalculating the phase for a shorter solution interval of 30 seconds. In this step we also determine the normalised scalar amplitude solutions for the polarisation calibrator. After applying all solutions again we solve for the crosshand delay (Kcross) and for the crosshand phase (Xf), accounting for any ionospheric changes. Kcross we solve for using a solution interval of 30 seconds to account for temporal variations. For the crosshand phase specifically we ensure a robust solution by solving in \SI{20}{MHz} bins instead of for each channel, and by correcting for phase ambiguities in the solutions. This is done by determining the mean phase of the solution and checking if any step across frequency shows a shift of more than 90 degrees. If that is the case a phase of 180 degrees is added to this step. Such shifts are usually caused by the ambiguity of the direction of the polarisation vector (no visible difference can be seen between a polarisation vector at 0 degrees and one rotated by 180 degrees). Therefore adding these 180 degrees potentially fixes the solution steps in frequency that are not in line with the surrounding solution phases. As these solutions are not modified during self-calibration of the target field, it is important to ensure that crosshand phase solutions behave as expected and that the applied solutions to the polarisation calibrator show no strong deviations. Fig.~\ref{fig:crosshand_phase} and Fig.~\ref{fig:vis_phase} show the crosshand solutions and the visibilities in the crosshands for the polarisation calibrator for one of our observations.


For the crosshand phase small variations between 10 and 20 degrees are a good indication that the solutions are not affected by RFI. For the visibilities, the phase against frequency plot reveals any remaining RFI or even bad calibration through deviations from mean value of zero. For the crosshand phase in our calibration we can see that the variations are less than 20 degrees. However, in the crosshands of the polarisation calibrator remaining RFI can still be seen, indicating that the flagging strategy and derivation of the calibration solutions has still room for improvement.

After finishing the polarisation calibration, we rerun the phase calibration of the secondary calibrator (G), now taking the crosshand phase and delay solutions into account. The secondary calibrator J$1150-0023$ is polarised at the level of a few percent, thus affecting its phase to a certain degree. Removing this effect on the gain phase solution by taking the polarisation into account may contribute to a cleaner phase solution applied to the target fields and could generate more robust initial conditions for the self-calibration round later.

\subsection{Target self-calibration and subtraction of M\,87}
After deriving the solution tables from the calibrators, we apply them to each of the four target fields before they are averaged using linear interpolation in both time and frequency. In detail these solution tables are the delay, phase and residual amplitude solutions of the gain calibrator, the amplitude, bandpass and leakage solutions of the bandpass calibrator, and the crosshand delay and crosshand phase solutions of the polarisation calibrator. Afterwards we average the target fields and apply ionospheric Faraday rotation corrections using again \texttt{Spinifex} and \texttt{DP3} for each individual scan of each target field. The contribution of the ionospheric RM is usually low for the frequency range of the L-band. We find this as well for the fields in our observations. The mean ionospheric RM across the fields is $\SI{-0.8}{rad/m^2}$. The ionospheric RM values for each field separately are reported in Table \ref{tab:qc_props_target}.

The self-calibration of the target fields is done using the self-calibration software \texttt{facetselfcal}. This software iteratively images the field to create a model column, derives antenna-based calibration solutions and applies them to the data, using \texttt{DP3} and in our case \texttt{WSClean} as an imager. We use \texttt{facetselfcal} for direction-independent calibration only, as no bright source is causing artefacts within the fields. In total, we run four rounds of calibration, starting with two rounds of scalar phase calibration followed by another two rounds of scalar complex gain calibration on the total intensity emission. The phase-only calibration runs are performed on short solution intervals of one minute, to correct for fast changing phase variations. To ensure stable solutions, we apply smoothness constraints of $\SI{100}{MHz}$ to these solutions. For the amplitude and phase solutions, calculated in the last two rounds, we increase the solution interval to 30 minutes, as we expect more slowly changing amplitudes. With a lower smoothness constraint of $\SI{50}{MHz}$ we ensure stable solutions for the amplitude. The amplitude solutions are additionally clipped for values above 1.5 and below 0.667. Throughout the self-calibration we restrict the solutions to scalar gains (i.e. identical solutions for the parallel hands) to preserve the relative calibration between HH and VV and therefore the amplitude of Stokes $Q$. We also use the \texttt{-forwidefield}-setting to be able to use the solutions for wide-field imaging. For the imaging of the data \texttt{WSClean} is used. As only the total intensity is imaged, we set the \texttt{-multiscale} parameter and image the data at a pixelscale of 2'' for an image with a size of 6000 pixels to deconvolve any potential bright sources that could affect the field.

After self-calibration we remove the effects caused by M\,87. By imaging the target fields phase shifted to M\,87, we are able to measure the flux of the contamination of M\,87 into each target field. If this flux is higher than $5\sigma$ of the surrounding noise, we use the \texttt{MODEL\_DATA} column produced during the imaging and subtract it from the data. This way we remove the residual effects of M\,87 on the target fields whenever necessary. For the fields regarded in this study, we used the simplest subtraction method of M\,87 where we do not subtract the field prior to phase shifting and imaging of M\,87. This is applicable for these fields as they are not strongly affected by M\,87. For fields closer to M\,87, however, the subtraction of the field before phase shifting or a self-calibration of M\,87 is likely necessary. Table \ref{tab:qc_props_target} lists some imaging quality metrics for each of the four target fields (flagged percentage, Ionospheric RM, noise before and after self-calibration and M\,87 subtraction and the dynamic range). 

\begin{table*}[ht]
\caption{Summary of imaging quality metrics for all target fields, including the flagged percentage, the noise before and after self-calibration and the final dynamic range.}
\label{tab:qc_props_target}     
\centering                      
\begin{tabular}{c| c c c c c c}    
\hline
\Tstrut
Target field & Flagged data & RM$_{\mathrm{ionosphere}}$ & $\mathrm{rms}_{\mathrm{before}}$ & $\mathrm{rms}_{\mathrm{after}}$ & $f = \frac{\mathrm{rms}_{\mathrm{before}}}{\mathrm{rms}_{\mathrm{after}}}$ & dynamic range\\
\Tstrut
 & [$\%$] & [rad/m$^2$] & [$\mu$Jy/beam] & [$\mu$Jy/beam] & & \\
\hline
\Tstrut
North-East & 49.8 & $-0.95$ & 10.72 & 8.52 & 1.26 & 7050\\
North-West & 51.7 & $-0.79$ & 10.15 & 8.68 & 1.17 & 10470\\
South-West & 51.7 & $-0.77$ & 12.41 & 10.54 & 1.18 & 17130\\
South-East & 55.9 & $-0.69$ & 10.80 & 10.01 & 1.08 & 9290\\
\hline 
\end{tabular}
\tablefoot{Column 1: Target field as oriented on sky; Column 2: percentage of data flagged after calibration; Column 3,4: MAD-based standard deviation of residual image before self-calibration and after self-calibration and M\,87 subtraction; Column 5: fraction of improvement in noise level; Column 6: dynamical range after M\,87 subtraction given as the ratio of peak emission to noise.}
\end{table*}

\subsection{Imaging, Mosaicking and RM synthesis}

In preparation for RM Synthesis, images in Stokes $I$, $Q$ and $U$ have been created using \texttt{WSClean}. We separated the imaging for the total intensity and the polarised emission, based on the different imaging requirements needed to optimise the images independently from each other. The images in total intensity are created by dividing the band into 12 ranges and producing 12 images. In total intensity, where at this depth sources are often resolved, we used multi-scale cleaning. Furthermore, the spectral slope of the detected radio emission can be made more robust against outliers by using a polynomial fit of order 3. This last option cannot be used in the imaging of Stokes $Q$ and $U$, as the Faraday rotation imposes strong spectral variation on the selected clean components. \texttt{WSClean} has settings optimised for imaging in polarisation, such as the \texttt{-join-polarization} and the \texttt{-fit-rm} parameter which were not used for the continuum imaging run. With the \texttt{-join-polarization} parameter, the peak finding during the cleaning is done in the combined Stokes $QU$ space instead of being done for each polarisation separately. In combination with \texttt{-join-frequency} and \texttt{-squared-channel-joining} it is possible to find peaks over the sum of channels in $Q^2 + U^2$. The \texttt{-fit-rm} parameter will fit each component to an RM and perform RM synthesis within the cleaning. With this the cleaning can become deeper and more accurate \citep{2017MNRAS.471..301O}.

To avoid bandpass depolarisation, the band width of each channel should be \SI{< 10}{MHz}. By setting the output channel number to 100 for the polarised imaging, each channel has a bandwidth of \SI{7.5}{MHz} resulting in a detectable Faraday depth of $\SI{790}{rad/m^2}$. To match the resolution across the bandwidth, a taper is applied and the restoring beam is forced to be circular with a size of $13''$. With this beam and taper, we are able to detect diffuse emission, even in polarisation, while still being able to resolve the structure within extended sources. The images have been created using a pixel scale of $2''$ and an image size of 6000 pixels. This image size ensures that bright sources in the first side-lobe are properly deconvolved.

As a final quality control of the calibration we compare the flux of the sources per target field to the ones in the Rapid ASKAP Continuum Survey at low frequencies (RACS-low, $\SI{888}{MHz}$). We perform this check after the full calibration sequence (self-calibration and subtraction of M\,87) has been applied and each pointing has been imaged as explained above. We use the MFS total intensity maps of each field and correct them for the primary beam at the reference frequency of $\SI{1284}{MHz}$, identical to the mean frequency of the image. The primary beam correction is done using the beam model by \cite{2020ApJ...888...61M}. We then create a catalogue of the sources contained in the primary beam corrected field using \texttt{pyBDSF} \citep{2015ascl.soft02007M}, with its standard settings for detection. The thus created catalogue is then filtered such that following conditions are fulfilled. The source needs to be within the primary beam determined with the model by \cite{2020ApJ...888...61M}, the source needs to have a flux at 5$\sigma$ above the determined uncertainty and the source needs to be at least 60 arcseconds away from any other source. By enforcing this we ensure that only isolated sources with a high SNR are detected within the primary beam. The same filters are applied to the RACS-low catalogue, before cross-matching them. The allowed maximum separation between the matched sources is set to 10 arcseconds, allowing for slight differences in position. For each source in the field matched to a RACS-low source we determine the flux ratio and calculate the mean flux ratio per field.    

With the per field comparison we are able to determine whether any issues occurred during calibration as the four fields were observed in three different observations and therefore calibrated separately. Across the four fields we find a mean flux ratio varying between $\frac{f_{\mathrm{RACS}}}{f_{\mathrm{MeerKAT}}} = 1.333$ and $1.325$. The expected flux ratio for the same source at $\SI{888}{MHz}$ and $\SI{1284}{MHz}$ is $\frac{f_{\mathrm{RACS}}}{f_{\mathrm{MeerKAT}}} \approx 1.343$ for an assumed spectral index of $\alpha = -0.8$, consistent with the expected spectral index for this frequency range. We calculate this using $\log(\frac{S_1}{S_2}) = \alpha \log(\frac{\nu_1}{\nu_2})$, where $S_1$ and $S_2$ are the fluxes of the sources for RACS-low and MeerKAT L-band respectively and $\nu_1$ and $\nu_2$ their frequencies. Since the pointings are overlapping, the same source appears in different fields and can therefore distort the observed mean flux ratio. To ensure that this is not the case, we create a catalogue of the sources across the full mosaic as well and cross-match it again with the RACS-low sources, using the same procedure as described above. Doing so we find 53 unique sources across the mosaic with a mean flux ratio of $1.334 \pm 0.024$, agreeing with the values for the individual fields. All values show no significant deviation from the expected flux ratio. With this we can confirm that our calibration returns an acceptable flux scale of the found sources. The field by field, and the mosaic flux ratio are given in Table \ref{tab:flux_comp_target}.

\begin{table}[ht]
\caption{Comparison of the flux ratio between RACS-low ($\SI{888}{MHz}$) and MeerKAT L-band ($\SI{1284}{MHz}$) for the sources in the primary beam corrected target fields.}
\label{tab:flux_comp_target}     
\centering                      
\begin{tabular}{c| c c c c c c}    
\hline
\Tstrut
Target field & \# of sources & $\frac{f_{\mathrm{RACS}}}{f_{\mathrm{MeerKAT}}}$ \\
\hline
\Tstrut
North-East & $17$ & $1.325 \pm 0.086$ \\
North-West & $20$ & $1.327 \pm 0.039$ \\
South-West & $31$ & $1.333 \pm 0.028$ \\
South-East & $23$ & $1.329 \pm 0.055$ \\
\hline
\Tstrut
Mosaic & $53$ & $1.334 \pm 0.024$\\
Theoretical & & 1.343\\
\hline 
\end{tabular}
\end{table}

Besides the Stokes I images mentioned above, we also image the visibility data at high resolution without tapering the weights, yielding an angular resolution $\approx 8''$. For the purpose of mosaicking, we enforce a circular restoring beam of $8''$ in this imaging step. We reduce the pixel scale to $1''$. These high-resolution total emission images are used for tracing finer structures in individual sources, like M\,49.

After imaging each target field separately, each channel of the target fields is combined separately, creating a mosaic for each channel. During that process the images are primary beam corrected. This was done using the software \texttt{MosaicQueen}. In the process of mosaicking the images have been cut at 50\% FWHM of the primary beam, reducing the effect of off-axis leakage occurring in the higher frequency bands at the edges of the primary beam \citep{2023AJ....165...78D}. In preparation for RM Synthesis the mosaicked channels in Stokes $Q$ and $U$ have been collected in cubes. Due to the high channel number, it is possible for most data within a channel to be flagged and therefore for the channel to be very noisy. The median noise in Stokes $Q$ and $U$ across the band is $\SI{95}{\mu Jy/beam}$, however, some channels show a mean noise of up to $\SI{1800}{\mu Jy/beam}$, deviating strongly from the median noise across the band. We exclude such channels from the cube by setting a limit of $\SI{500}{\micro Jy/beam}$, at cube creation for the noise in both, the Stokes $Q$ and $U$ images. This corresponds to a cut at approximately 5 times the median noise, ensuring that  only channels are removed which could impact the quality of the data during RM synthesis even when weighted. The more subtle differences in noise are accounted for via noise weighting during RM Synthesis. 

We are then able to determine the polarisation properties of the field by applying the RM Synthesis technique developed by \cite{1966MNRAS.133...67B, 2005A&A...441.1217B}. More precisely, we employ the \texttt{RMSynth3D} routine of \texttt{RM-Tools} to produce the Faraday dispersion function (FDF) and the rotation measure spread function (RMSF) for the field. We limit the Faraday peak-fitting between -300 and $\SI{300}{rad/\meter^2}$ with a recommended sampling of 10 per full width half maximum (FWHM), corresponding to a step size of roughly $\SI{4}{rad/\meter^2}$. As each channel in the cube may be flagged differently and hence the noise within each image varies, we weight each channel by its inverse noise squared $\frac{1}{\sigma_{QU}^2}$. This is implemented by setting the \texttt{-w} flag to `variance' and providing a noise file. The noise file used for weighting the channels is given by $\sigma_{QU} = \sqrt{\sigma_Q^2 + \sigma_U^2}$ calculated for each channel. Including this improves the accuracy of the RM calculation at the cost of increasing the width of the RMSF \citep{2009IAUS..259..591H}. 

Based on our observational parameters, we can calculate the maximum probed Faraday depth, the resolution in Faraday depth and the maximal resolvable scale as follows:
\begin{equation}
    ||\phi_{\rm max}|| \approx \frac{\sqrt{3}}{\delta \lambda^2} \approx \SI{2336}{rad/\meter^2}
\end{equation}
\begin{equation*}
    \delta\phi \approx \frac{2\sqrt{3}}{\Delta \lambda^2} \approx \SI{41}{rad/\meter^2}
\end{equation*}
\begin{equation*}
    \mathrm{max-scale} \approx \frac{\pi}{ \lambda^2_{\mathrm{min}}} \approx \SI{95}{rad/\meter^2} .
\end{equation*}
Here $\delta\lambda$ is the channel width, $\Delta\lambda$ is the total width of the $\lambda^2$ distribution, and $\lambda_{\mathrm{min}}$ is the shortest wavelength.

Using this FDF cube we calculate the final maps containing the polarisation properties. For the polarised emission map we assume that all sources are Faraday simple and therefore have only a single peak in Faraday depth, with its peak value representing the polarised intensity. Similarly, we derive the RM map of the field by determining the position of the maximum Faraday peaks, assuming Faraday-simple sources.

To determine the noise map for the polarised intensity image, which is also needed for the calculation of the RM map, we cannot use the $\sigma_{QU}$ used for the RM clean threshold, as this value will not account for any spatial variations. Instead we calculate a noise map from the FDF cube directly. To do this we choose channels at high Faraday depth (here: $|\phi| \geq \SI{100}{rad/m^2}$), where we do not expect a signal from the sources in Stokes $Q$ and $U$. We compute the MAD-based standard deviation of these channels at each pixel in both $Q$ and $U$ and apply a Gaussian filter with a width of 20 pixels to ensure that the noise varies smoothly across the field. Thus, noise variations across the image are taken into account and the inclusion of false source detections, especially in the outer regions, is reduced. The polarised noise map is then given by $\frac{\sigma_Q + \sigma_U}{2}$. This noise map, which we call $\sigma_P$ throughout the paper, is used for correcting the Ricean bias in the polarised emission maps, as the threshold for polarised source detection and to determine the uncertainty of the RM map. The latter is given by
\begin{equation}
    \mathrm{RM_{err}} = \frac{\mathrm{\delta \phi}}{2\frac{P}{\sigma_P}}, 
\end{equation}
where $\mathrm{\delta \phi}$ is the FWHM of the RM spread function, P is the polarised emission and $\sigma_{P}$ is the noise of the polarised map \citep{2013A&A...552A..58S}.

The FDF cube used for the polarisation and RM map creation is still convolved with the RMSF and therefore not appropriate for the determination of Faraday simplicity of the sources in the field. For the analysis of the Faraday spectrum of the sources, especially the centre and outer lobe analysis of M\,49, the FDF has been cleaned, using the \texttt{RM-Tools} task \texttt{RMClean3D} which, similar to aperture synthesis imaging, can deconvolve the FDF \citep{2009IAUS..259..591H}. We calculate the theoretical noise in polarised emission using:
\begin{equation}
    \sigma_{P,\mathrm{theo}} = \frac{1}{\sqrt{\sum \limits_{i} (\frac{1}{\sigma_i})^2}} \approx \SI{14.71}{\mu Jy},
\end{equation}
where $\sigma_i$ is the noise per channel. The FDF was cleaned down to $5\sigma_{P,\mathrm{theo}}$. This value is the global noise across the whole of the mosaic. As the mosaic consists of regions with overlapping fields and regions with single fields, the noise will vary locally, which is not reflected in $\sigma_{P,\mathrm{theo}}$. Therefore, the outer regions will be deconvolved below the local $5\sigma$, while the innermost regions will be cleaned to a higher $\sigma$-threshold.

\subsection{Off-axis leakage above 1.4\,GHz}
\label{subsec:leakage}
As has been shown by \cite{LinPolCal2024Memo} the effect of off-axis leakage can be substantial at frequencies above $\SI{1.4}{GHz}$, increasing the further away from the phase centre the measurement is made. Off-axis leakage can lead to an increase in the observed polarised emission as well as spurious emission, which is not related to any physical effect but an artefact from the telescope beam. One way to mitigate this effect is by restricting the analysis of polarisation to the frequency band up to $\SI{1.4}{GHz}$, as has been done in other studies \citep[see for example][]{2025A&A...694A.125L, 2024MNRAS.528.2511T}. In this study we cut the images before mosaicking at 50\% FWHM of the primary beam per frequency channel, generating different sized images per channel restricting the data at high frequencies to regions closer to the phase centre. Additionally we mosaic the observed pointings with a phase centre distance of 0.58°. Using these methods the effect of off-axis leakage should be reduced across the polarised image.

\begin{figure}[ht]
     \centering
     \includegraphics[width=0.9\columnwidth]{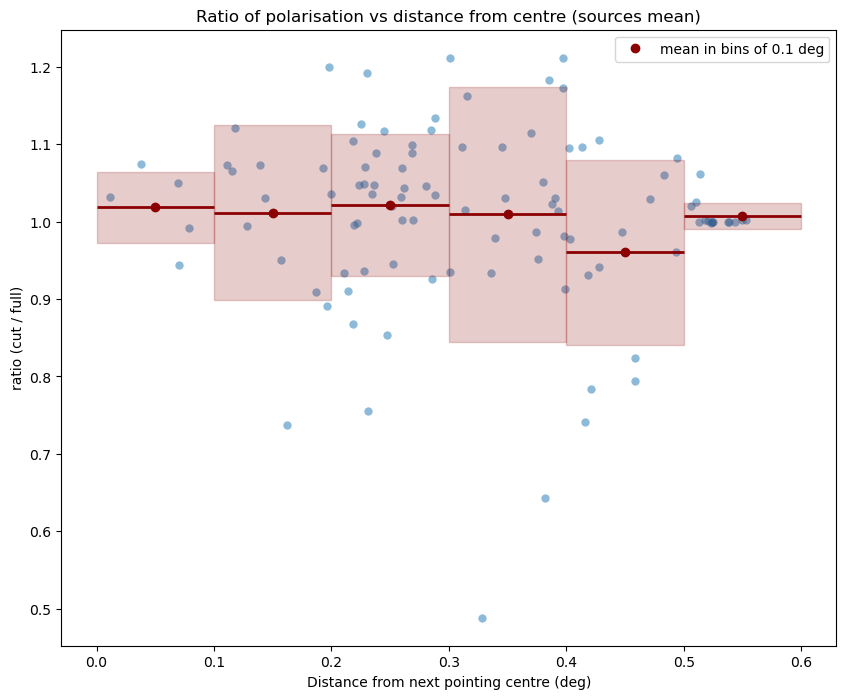}
     \caption{Off-axis leakage test in polarised emission. Ratio of the polarised intensity per source measured in the $0.9 - \SI{1.4}{GHz}$ map to that measured in the full $0.9 - \SI{1.65}{GHz}$ map, as a function of distance from the nearest pointing centre. The red points and lines mark the mean values per distance bin. The shaded red area shows the standard deviation per bin.}
     \label{fig:pol_ratio}
\end{figure}

\begin{figure}[ht]
     \centering
     \includegraphics[width=0.9\columnwidth]{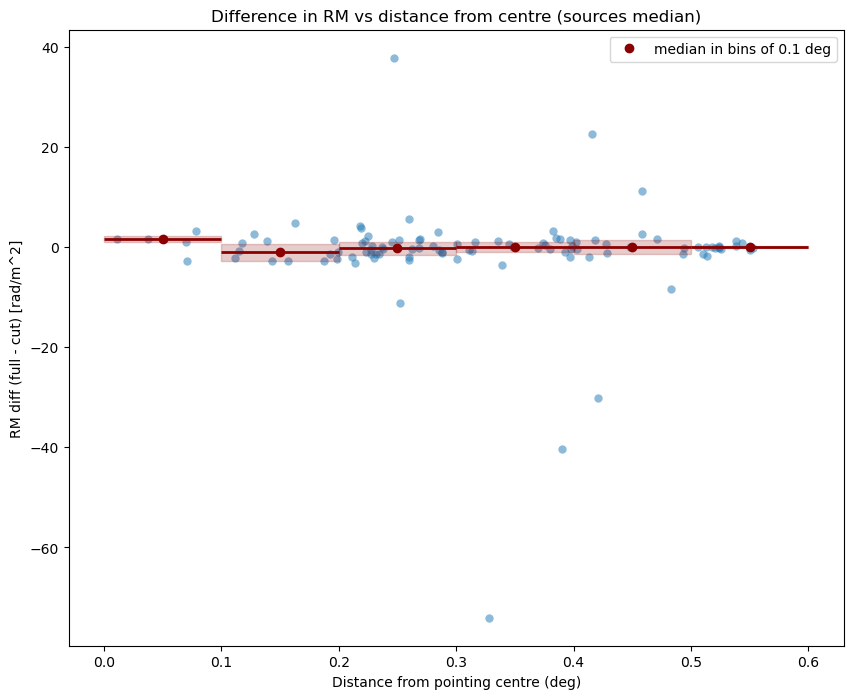}
     \caption{Off-axis leakage test in RM. The difference between the Rotation Measure for the case where the frequency is cut at \SI{1.4}{GHz} and for the full frequency band is plotted against the distance from the nearest pointing centre. The red points and lines mark the median values per distance bin. The shaded red area shows the MAD per bin.}
     \label{fig:RM_diff}
\end{figure}

To test whether this is the case, we recreate the polarised intensity map and the RM map using a Stokes $Q$ and $U$ cube only containing data below $\SI{1.4}{GHz}$. Using the same mask as for the full frequency band data, we can compare the polarised intensity and RM per source with each other. In Figure \ref{fig:pol_ratio} we calculate the mean polarised emission across the source for both the cut and the full map and calculate their ratio $\frac{P_{\mathrm{cut}}}{P_{\mathrm{full}}}$. If leakage contributed more to the full-band map than to the cut one, a continuous decrease in the fraction should be visible in relation to the distance of the pointing centre. We plot against the distance to the nearest pointing centre within the mosaic and calculate the mean values in bins of 0.1°.
For the mean values per bin, we do not see any significant decline towards larger distances from the phase centre. The ratio per bin varies between 0.96 and 1.02 with the standard deviation changing between 0.04 and 0.16. No clear trend with distance is apparent. A similar behaviour is seen in the comparison of the GRM uncorrected RM values. Similar as in our analysis for the paper we calculate the polarised emission weighted mean of the RM per source for both the full frequency band data and the data cut at $\SI{1.4}{GHz}$. We then plot the difference $\mathrm{RM_{full}} - \mathrm{RM_{cut}}$ against the pointing centre distance again. For the values per bin we calculate this time the median and the MAD, to check for any relation to the distance. This is shown in Figure \ref{fig:RM_diff}.
Again no increasing difference in RM can be seen with increasing distance from the pointing centre. The bin median of the RM difference only goes as high as $\SI{1.6}{rad/m^2}$, with the MAD being similarly low (the maximum value is in the second bin at $\SI{1.7}{rad/m^2}$). 

Based on these tests we conclude that the off-axis leakage is indeed mitigated by the primary beam restriction and the mosaicking pattern, to a level comparable to that obtained by removing the data above $\SI{1.4}{GHz}$. We therefore use RM Synthesis and do the polarisation analysis on the data in the frequency range of $\SI{900}{MHz}−\SI{1.65}{GHz}$.

\section{RRM Excess along multiple directions around M\,49}

\begin{figure}[ht]
     \centering
     \includegraphics[width=0.9\columnwidth]{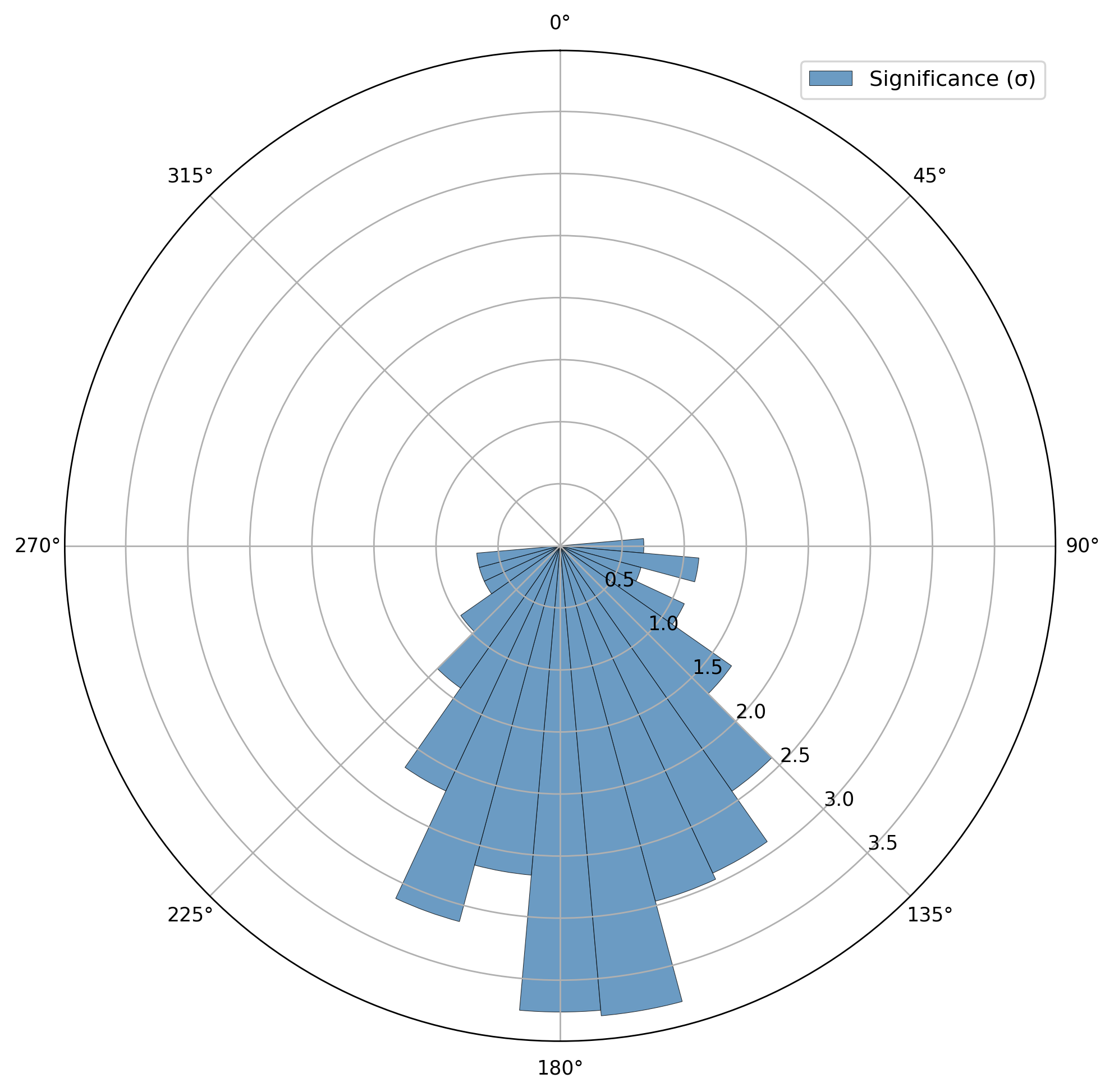}
     \caption{Polar plot of the excess-scatter significance for each direction around M\,49 in steps of 10°. The significance is given in units of $\sigma$. The direction along which the highest significance is measured is at 170°, close to the N-S direction.}
     \label{fig:dir_scatter}
\end{figure}

\begin{table*}[ht]
\caption{Directional RRM scatter and excess scatter around M\,49 in steps of 10°.}             
\label{tab:dir_scatter}      
\centering          
\begin{tabular}{c c c c c c c c c c c }   
\hline
\Tstrut                   
Dir & $N_A$ & $N_B$ & $M(\mathrm{RRM}_A)$ & $M(\mathrm{RRM}_B)$ & $\sigma_A$ & $\sigma_B$ & $\sigma_{\mathrm{excess},AB}$ & sig$_{AB}$ & $\sigma_{\mathrm{excess},BA}$ & sig$_{BA}$\\
\Tstrut 
 & & & [$\mathrm{rad\,m}^{-2}$] & [$\mathrm{rad\,m}^{-2}$] & [$\mathrm{rad\,m}^{-2}$] & [$\mathrm{rad\,m}^{-2}$] & [$\mathrm{rad\,m}^{-2}$] & [$\sigma$] & [$\mathrm{rad\,m}^{-2}$] & [$\sigma$]\\
\hline
\Tstrut 
0 & 43 & 58 & $-4.51$ & $-8.58$ & $8.5 \pm 2.2$ & $14.4 \pm 2.1$ & \dots & \dots & $11.6 \pm 3.1$ & 3.8 \\
10 & 38 & 63 & $-4.11$ & $-8.18$ & $9.2 \pm 2.3$ & $13.3 \pm 2.0$ & \dots & \dots & $9.5 \pm 3.6$ & 2.7 \\
20 & 36 & 65 & $-4.11$ & $-8.74$ & $8.7 \pm 2.3$ & $13.5 \pm 2.0$ & \dots & \dots & $10.2 \pm 3.3$ & 3.1 \\
30 & 37 & 64 & $-3.16$ & $-8.86$ & $9.4 \pm 2.4$ & $12.8 \pm 2.0$ & \dots & \dots & $8.5 \pm 3.9$ & 2.2 \\
40 & 38 & 63 & $-4.11$ & $-8.03$ & $10.3 \pm 2.3$ & $12.5 \pm 2.0$ & \dots & \dots & $7.0 \pm 5.0$ & 1.4 \\
50 & 39 & 62 & $-3.72$ & $-8.11$ & $10.7 \pm 2.4$ & $12.2 \pm 2.0$ & \dots & \dots & $5.9 \pm 6.0$ & 1.0 \\
60 & 40 & 61 & $-4.65$ & $-7.38$ & $11.5 \pm 2.3$ & $12.0 \pm 2.0$ & \dots & \dots & $3.4 \pm 5.1$ & 0.7 \\
70 & 38 & 63 & $-5.95$ & $-6.48$ & $11.8 \pm 2.4$ & $11.9 \pm 1.9$ & \dots & \dots & $1.4 \pm 2.1$ & 0.7 \\
80 & 37 & 64 & $-6.98$ & $-6.31$ & $11.4 \pm 2.4$ & $12.1 \pm 1.9$ & \dots & \dots & $4.0 \pm 6.0$ & 0.7 \\
90 & 41 & 60 & $-9.35$ & $-4.86$ & $12.0 \pm 2.4$ & $11.4 \pm 2.0$ & $3.7 \pm 5.5$ & 0.7 & \dots & \dots \\
100 & 42 & 59 & $-8.49$ & $-4.91$ & $12.8 \pm 2.4$ & $11.0 \pm 2.0$ & $6.5 \pm 5.8$ & 1.1 & \dots & \dots \\
110 & 43 & 58 & $-9.35$ & $-4.69$ & $12.0 \pm 2.4$ & $11.4 \pm 1.9$ & $3.8 \pm 5.7$ & 0.7 & \dots & \dots \\
120 & 46 & 55 & $-9.16$ & $-4.86$ & $12.7 \pm 2.3$ & $11.0 \pm 2.0$ & $6.4 \pm 5.8$ & 1.1 & \dots & \dots \\
130 & 50 & 51 & $-9.16$ & $-4.86$ & $13.2 \pm 2.2$ & $10.6 \pm 2.1$ & $7.9 \pm 4.7$ & 1.7 & \dots & \dots \\
140 & 54 & 47 & $-8.30$ & $-4.91$ & $13.7 \pm 2.2$ & $9.9 \pm 2.1$ & $9.3 \pm 3.9$ & 2.4 & \dots & \dots \\
150 & 56 & 45 & $-7.63$ & $-4.96$ & $14.0 \pm 2.2$ & $9.5 \pm 2.1$ & $10.3 \pm 3.5$ & 2.9 & \dots & \dots \\
160 & 57 & 44 & $-8.30$ & $-4.86$ & $14.0 \pm 2.1$ & $9.4 \pm 2.1$ & $10.3 \pm 3.5$ & 3.0 & \dots & \dots \\
170 & 56 & 45 & $-8.98$ & $-4.69$ & $14.8 \pm 2.2$ & $8.7 \pm 2.1$ & $11.9 \pm 3.1$ & 3.8 & \dots & \dots \\
\hline                  
\end{tabular}
\tablefoot{Column 1: Direction of scatter through N-W; Columns 2,3: Number of sources in the two regions; Columns 4,5: median RRM in the two regions; Columns 6,7: RRM scatter in both regions; Columns 8,9: excess scatter and its significance of region A over B; Columns 10,11: excess scatter and its significance of region B over A.}
\end{table*}

\end{appendix}

\end{document}